\documentclass[journal,draftclsnofoot,onecolumn,12pt]{IEEEtran}
\usepackage{graphicx}
\usepackage{subfigure}
\usepackage{multirow}
\usepackage{array}
\newcolumntype{P}[1]{>{\centering\arraybackslash}p{#1}}
\newcolumntype{M}[1]{>{\centering\arraybackslash}m{#1}}

\usepackage{amsthm,amssymb,amsmath,graphicx,multirow,color,amsfonts}%
\usepackage[update,prepend]{epstopdf}
\usepackage{multirow}
\usepackage[latin1]{inputenc}
\usepackage{tikz}
\usepackage{bbm} 
\usepackage{pdfpages}
\usepackage{multirow}
\usepackage{subfig}
\usepackage{comment}
 \usepackage{graphicx}
\usepackage{multicol}
\usepackage{cite}

\usepackage[justification=centering]{caption}
\usepackage{textcomp}
\usepackage{psfrag}
\usepackage{arydshln}
\usepackage{url}
\usepackage{soul}
\usepackage{graphicx,color}
\usepackage[nolist]{acronym}
\usepackage{algorithm,algorithmic} 

\usepackage{mathtools,lipsum}
\usepackage{cuted}
\usepackage{amsmath}
\newtheorem{proposition}{Proposition} 
\usepackage{mathrsfs}
\usepackage{color}

\usepackage[capitalise]{cleveref}
\Crefname{equation}{Eq.\!}{Eqs.\!}
\Crefname{figure}{Fig.\!}{Figs.\!}
\Crefname{tabular}{Tab.\!}{Tabs.\!}
\Crefname{section}{Section\!}{Sections.\!}

\def\nb0{{\mathbf{0}}}
\def\nb1{{\mathbf{1}}}

\newtheorem{lemma}{Lemma}

\newtheorem{definition}{Definition}

\begin{document}
\graphicspath{{./Figures/}}
	\begin{acronym}

\acro{PSO}{ particles swarm optimization}

\acro{RF}{radio frequency}
\acro{THz}{terahertz}
\acro{HRTRN}{hybrid RF and THz relay network}

\acro{5G-NR}{5G New Radio}
\acro{3GPP}{3rd Generation Partnership Project}
\acro{ABS}{aerial base station}
\acro{AC}{address coding}
\acro{ACF}{autocorrelation function}
\acro{ACR}{autocorrelation receiver}
\acro{ADC}{analog-to-digital converter}
\acrodef{aic}[AIC]{Analog-to-Information Converter}     
\acro{AIC}[AIC]{Akaike information criterion}
\acro{aric}[ARIC]{asymmetric restricted isometry constant}
\acro{arip}[ARIP]{asymmetric restricted isometry property}

\acro{ARQ}{Automatic Repeat Request}
\acro{AUB}{asymptotic union bound}
\acrodef{awgn}[AWGN]{Additive White Gaussian Noise}     
\acro{AWGN}{additive white Gaussian noise}

\acro{APSK}[PSK]{asymmetric PSK} 

\acro{waric}[AWRICs]{asymmetric weak restricted isometry constants}
\acro{warip}[AWRIP]{asymmetric weak restricted isometry property}
\acro{BCH}{Bose, Chaudhuri, and Hocquenghem}        
\acro{BCHC}[BCHSC]{BCH based source coding}
\acro{BEP}{bit error probability}
\acro{BFC}{block fading channel}
\acro{BG}[BG]{Bernoulli-Gaussian}
\acro{BGG}{Bernoulli-Generalized Gaussian}
\acro{BPAM}{binary pulse amplitude modulation}
\acro{BPDN}{Basis Pursuit Denoising}
\acro{BPPM}{binary pulse position modulation}
\acro{BPSK}{Binary Phase Shift Keying}
\acro{BPZF}{bandpass zonal filter}
\acro{BSC}{binary symmetric channels}              
\acro{BU}[BU]{Bernoulli-uniform}
\acro{BER}{bit error rate}
\acro{BS}{base station}
\acro{BW}{BandWidth}
\acro{BLLL}{ binary log-linear learning }

\acro{CP}{Cyclic Prefix}
\acrodef{cdf}[CDF]{cumulative distribution function}   
\acro{CDF}{Cumulative Distribution Function}
\acrodef{c.d.f.}[CDF]{cumulative distribution function}
\acro{CCDF}{complementary cumulative distribution function}
\acrodef{ccdf}[CCDF]{complementary CDF}               
\acrodef{c.c.d.f.}[CCDF]{complementary cumulative distribution function}
\acro{CD}{cooperative diversity}

\acro{CDMA}{Code Division Multiple Access}
\acro{ch.f.}{characteristic function}
\acro{CIR}{channel impulse response}
\acro{cosamp}[CoSaMP]{compressive sampling matching pursuit}
\acro{CR}{cognitive radio}
\acro{cs}[CS]{compressed sensing}                   
\acrodef{cscapital}[CS]{Compressed sensing} 
\acrodef{CS}[CS]{compressed sensing}
\acro{CSI}{channel state information}
\acro{CCSDS}{consultative committee for space data systems}
\acro{CC}{convolutional coding}
\acro{Covid19}[COVID-19]{Coronavirus disease}

\acro{DAA}{detect and avoid}
\acro{DAB}{digital audio broadcasting}
\acro{DCT}{discrete cosine transform}
\acro{dft}[DFT]{discrete Fourier transform}
\acro{DR}{distortion-rate}
\acro{DS}{direct sequence}
\acro{DS-SS}{direct-sequence spread-spectrum}
\acro{DTR}{differential transmitted-reference}
\acro{DVB-H}{digital video broadcasting\,--\,handheld}
\acro{DVB-T}{digital video broadcasting\,--\,terrestrial}
\acro{DL}{DownLink}
\acro{DSSS}{Direct Sequence Spread Spectrum}
\acro{DFT-s-OFDM}{Discrete Fourier Transform-spread-Orthogonal Frequency Division Multiplexing}
\acro{DAS}{Distributed Antenna System}
\acro{DNA}{DeoxyriboNucleic Acid}

\acro{EC}{European Commission}
\acro{EED}[EED]{exact eigenvalues distribution}
\acro{EIRP}{Equivalent Isotropically Radiated Power}
\acro{ELP}{equivalent low-pass}
\acro{eMBB}{Enhanced Mobile Broadband}
\acro{EMF}{ElectroMagnetic Field}
\acro{EU}{European union}
\acro{EI}{Exposure Index}
\acro{eICIC}{enhanced Inter-Cell Interference Coordination}

\acro{FC}[FC]{fusion center}
\acro{FCC}{Federal Communications Commission}
\acro{FEC}{forward error correction}
\acro{FFT}{fast Fourier transform}
\acro{FH}{frequency-hopping}
\acro{FH-SS}{frequency-hopping spread-spectrum}
\acrodef{FS}{Frame synchronization}
\acro{FSsmall}[FS]{frame synchronization}  
\acro{FDMA}{Frequency Division Multiple Access}

\acro{GA}{Gaussian approximation}
\acro{GF}{Galois field }
\acro{GG}{Generalized-Gaussian}
\acro{GIC}[GIC]{generalized information criterion}
\acro{GLRT}{generalized likelihood ratio test}
\acro{GPS}{Global Positioning System}
\acro{GMSK}{Gaussian Minimum Shift Keying}
\acro{GSMA}{Global System for Mobile communications Association}
\acro{GS}{ground station}
\acro{GMG}{ Grid-connected MicroGeneration}

\acro{HAP}{high altitude platform}
\acro{HetNet}{Heterogeneous network}

\acro{IDR}{information distortion-rate}
\acro{IFFT}{inverse fast Fourier transform}
\acro{iht}[IHT]{iterative hard thresholding}
\acro{i.i.d.}{independent, identically distributed}
\acro{IoT}{Internet of Things}                      
\acro{IR}{impulse radio}
\acro{lric}[LRIC]{lower restricted isometry constant}
\acro{lrict}[LRICt]{lower restricted isometry constant threshold}
\acro{ISI}{intersymbol interference}
\acro{ITU}{International Telecommunication Union}
\acro{ICNIRP}{International Commission on Non-Ionizing Radiation Protection}
\acro{IEEE}{Institute of Electrical and Electronics Engineers}
\acro{ICES}{IEEE international committee on electromagnetic safety}
\acro{IEC}{International Electrotechnical Commission}
\acro{IARC}{International Agency on Research on Cancer}
\acro{IS-95}{Interim Standard 95}

\acro{KPI}{Key Performance Indicator}

\acro{LEO}{low earth orbit}
\acro{LF}{likelihood function}
\acro{LLF}{log-likelihood function}
\acro{LLR}{log-likelihood ratio}
\acro{LLRT}{log-likelihood ratio test}
\acro{LoS}{Line-of-Sight}
\acro{LRT}{likelihood ratio test}
\acro{wlric}[LWRIC]{lower weak restricted isometry constant}
\acro{wlrict}[LWRICt]{LWRIC threshold}
\acro{LPWAN}{Low Power Wide Area Network}
\acro{LoRaWAN}{Low power long Range Wide Area Network}
\acro{NLoS}{Non-Line-of-Sight}
\acro{LiFi}[Li-Fi]{light-fidelity}
 \acro{LED}{light emitting diode}
 \acro{LABS}{LoS transmission with each ABS}
 \acro{NLABS}{NLoS transmission with each ABS}

\acro{MB}{multiband}
\acro{MC}{macro cell}
\acro{MDS}{mixed distributed source}
\acro{MF}{matched filter}
\acro{m.g.f.}{moment generating function}
\acro{MI}{mutual information}
\acro{MIMO}{Multiple-Input Multiple-Output}
\acro{MISO}{multiple-input single-output}
\acrodef{maxs}[MJSO]{maximum joint support cardinality}                       
\acro{ML}[ML]{maximum likelihood}
\acro{MMSE}{minimum mean-square error}
\acro{MMV}{multiple measurement vectors}
\acrodef{MOS}{model order selection}
\acro{M-PSK}[${M}$-PSK]{$M$-ary phase shift keying}                       
\acro{M-APSK}[${M}$-PSK]{$M$-ary asymmetric PSK} 
\acro{MP}{ multi-period}
\acro{MINLP}{mixed integer non-linear programming}

\acro{M-QAM}[$M$-QAM]{$M$-ary quadrature amplitude modulation}
\acro{MRC}{maximal ratio combiner}                  
\acro{maxs}[MSO]{maximum sparsity order}                                      
\acro{M2M}{Machine-to-Machine}                                                
\acro{MUI}{multi-user interference}
\acro{mMTC}{massive Machine Type Communications}      
\acro{mm-Wave}{millimeter-wave}
\acro{MP}{mobile phone}
\acro{MPE}{maximum permissible exposure}
\acro{MAC}{media access control}
\acro{NB}{narrowband}
\acro{NBI}{narrowband interference}
\acro{NLA}{nonlinear sparse approximation}
\acro{NLOS}{Non-Line of Sight}
\acro{NTIA}{National Telecommunications and Information Administration}
\acro{NTP}{National Toxicology Program}
\acro{NHS}{National Health Service}

\acro{LOS}{Line of Sight}

\acro{OC}{optimum combining}                             
\acro{OC}{optimum combining}
\acro{ODE}{operational distortion-energy}
\acro{ODR}{operational distortion-rate}
\acro{OFDM}{Orthogonal Frequency-Division Multiplexing}
\acro{omp}[OMP]{orthogonal matching pursuit}
\acro{OSMP}[OSMP]{orthogonal subspace matching pursuit}
\acro{OQAM}{offset quadrature amplitude modulation}
\acro{OQPSK}{offset QPSK}
\acro{OFDMA}{Orthogonal Frequency-division Multiple Access}
\acro{OPEX}{Operating Expenditures}
\acro{OQPSK/PM}{OQPSK with phase modulation}

\acro{PAM}{pulse amplitude modulation}
\acro{PAR}{peak-to-average ratio}
\acrodef{pdf}[PDF]{probability density function}                      
\acro{PDF}{probability density function}
\acrodef{p.d.f.}[PDF]{probability distribution function}
\acro{PDP}{power dispersion profile}
\acro{PMF}{probability mass function}                             
\acrodef{p.m.f.}[PMF]{probability mass function}
\acro{PN}{pseudo-noise}
\acro{PPM}{pulse position modulation}
\acro{PRake}{Partial Rake}
\acro{PSD}{power spectral density}
\acro{PSEP}{pairwise synchronization error probability}
\acro{PSK}{phase shift keying}
\acro{PD}{power density}
\acro{8-PSK}[$8$-PSK]{$8$-phase shift keying}
\acro{PPP}{Poisson point process}
\acro{PCP}{Poisson cluster process}
 
\acro{FSK}{Frequency Shift Keying}

\acro{QAM}{Quadrature Amplitude Modulation}
\acro{QPSK}{Quadrature Phase Shift Keying}
\acro{OQPSK/PM}{OQPSK with phase modulator }

\acro{RD}[RD]{raw data}
\acro{RDL}{"random data limit"}
\acro{ric}[RIC]{restricted isometry constant}
\acro{rict}[RICt]{restricted isometry constant threshold}
\acro{rip}[RIP]{restricted isometry property}
\acro{ROC}{receiver operating characteristic}
\acro{rq}[RQ]{Raleigh quotient}
\acro{RS}[RS]{Reed-Solomon}
\acro{RSC}[RSSC]{RS based source coding}
\acro{r.v.}{random variable}                               
\acro{R.V.}{random vector}
\acro{RMS}{root mean square}
\acro{RFR}{radiofrequency radiation}
\acro{RIS}{Reconfigurable Intelligent Surface}
\acro{RNA}{RiboNucleic Acid}
\acro{RRM}{Radio Resource Management}
\acro{RUE}{reference user equipments}
\acro{RAT}{radio access technology}
\acro{RB}{resource block}

\acro{SA}[SA-Music]{subspace-augmented MUSIC with OSMP}
\acro{SC}{small cell}
\acro{SCBSES}[SCBSES]{Source Compression Based Syndrome Encoding Scheme}
\acro{SCM}{sample covariance matrix}
\acro{SEP}{symbol error probability}
\acro{SG}[SG]{sparse-land Gaussian model}
\acro{SIMO}{single-input multiple-output}
\acro{SINR}{signal-to-interference plus noise ratio}
\acro{SIR}{signal-to-interference ratio}
\acro{SISO}{Single-Input Single-Output}
\acro{SMV}{single measurement vector}
\acro{SNR}[\textrm{SNR}]{signal-to-noise ratio} 
\acro{sp}[SP]{subspace pursuit}
\acro{SS}{spread spectrum}
\acro{SW}{sync word}
\acro{SAR}{specific absorption rate}
\acro{SSB}{synchronization signal block}
\acro{SR}{shrink and realign}

\acro{tUAV}{tethered Unmanned Aerial Vehicle}
\acro{TBS}{terrestrial base station}

\acro{uUAV}{untethered Unmanned Aerial Vehicle}
\acro{PDF}{probability density functions}

\acro{PL}{path-loss}

\acro{TH}{time-hopping}
\acro{ToA}{time-of-arrival}
\acro{TR}{transmitted-reference}
\acro{TW}{Tracy-Widom}
\acro{TWDT}{TW Distribution Tail}
\acro{TCM}{trellis coded modulation}
\acro{TDD}{Time-Division Duplexing}
\acro{TDMA}{Time Division Multiple Access}
\acro{Tx}{average transmit}

\acro{UAV}{Unmanned Aerial Vehicle}
\acro{uric}[URIC]{upper restricted isometry constant}
\acro{urict}[URICt]{upper restricted isometry constant threshold}
\acro{UWB}{ultrawide band}
\acro{UWBcap}[UWB]{Ultrawide band}   
\acro{URLLC}{Ultra Reliable Low Latency Communications}
         
\acro{wuric}[UWRIC]{upper weak restricted isometry constant}
\acro{wurict}[UWRICt]{UWRIC threshold}                
\acro{UE}{User Equipment}
\acro{UL}{UpLink}

\acro{WiM}[WiM]{weigh-in-motion}
\acro{WLAN}{wireless local area network}
\acro{wm}[WM]{Wishart matrix}                               
\acroplural{wm}[WM]{Wishart matrices}
\acro{WMAN}{wireless metropolitan area network}
\acro{WPAN}{wireless personal area network}
\acro{wric}[WRIC]{weak restricted isometry constant}
\acro{wrict}[WRICt]{weak restricted isometry constant thresholds}
\acro{wrip}[WRIP]{weak restricted isometry property}
\acro{WSN}{wireless sensor network}                        
\acro{WSS}{Wide-Sense Stationary}
\acro{WHO}{World Health Organization}
\acro{Wi-Fi}{Wireless Fidelity}

\acro{sss}[SpaSoSEnc]{sparse source syndrome encoding}

\acro{VLC}{Visible Light Communication}
\acro{VPN}{Virtual Private Network} 
\acro{RF}{Radio Frequency}
\acro{FSO}{Free Space Optics}
\acro{IoST}{Internet of Space Things}

\acro{GSM}{Global System for Mobile Communications}
\acro{2G}{Second-generation cellular network}
\acro{3G}{Third-generation cellular network}
\acro{4G}{Fourth-generation cellular network}
\acro{5G}{Fifth-generation cellular network}	
\acro{gNB}{next-generation Node-B Base Station}
\acro{NR}{New Radio}
\acro{UMTS}{Universal Mobile Telecommunications Service}
\acro{LTE}{Long Term Evolution}

\acro{QoS}{Quality of Service}
\end{acronym}
	
\newcommand{\SAR} {\mathrm{SAR}}
\newcommand{\WBSAR} {\mathrm{SAR}_{\mathsf{WB}}}
\newcommand{\gSAR} {\mathrm{SAR}_{10\si{\gram}}}
\newcommand{\Sab} {S_{\mathsf{ab}}}
\newcommand{\Eavg} {E_{\mathsf{avg}}}
\newcommand{\ft}{f_{\textsf{th}}}
\newcommand{\alphatf}{\alpha_{24}}

\title{
Terrain-Based UAV Deployment: \\Providing Coverage for Outdoor Users
}
\author{
Zhengying~Lou, Ruibo~Wang, Baha~Eddine~Youcef~Belmekki,~\IEEEmembership{Member,~IEEE,} Mustafa~A.~Kishk, {\em Member, IEEE}, and Mohamed-Slim~Alouini, {\em Fellow, IEEE}
\thanks{Zhengying Lou, Ruibo Wang, Baha Eddine Youcef Belmekki and Mohamed-Slim Alouini are with King Abdullah University of Science and Technology (KAUST), CEMSE division, Thuwal 23955-6900, Saudi Arabia. Mustafa A. Kishk is with the Department of Electronic Engineering, Maynooth University, Maynooth, W23 F2H6, Ireland. (e-mail: zhengying.lou@kaust.edu.sa; ruibo.wang@kaust.edu.sa; bahaeddine.belmekki@kaust.edu.sa; mustafa.kishk @mu.ie; slim.alouini@kaust.edu.sa).}
}
\maketitle

\vspace{-1.2cm}

\begin{abstract}
Deploying unmanned aerial vehicle (UAV) networks to provide coverage for outdoor users has attracted great attention during the last decade.
However, outdoor coverage is challenging due to the high mobility of crowds and the diverse terrain configurations causing building blockage.
Most studies use stochastic channel models to characterize the impact of building blockage on user performance and do not take into account terrain information.
On the other hand, real-time search methods use terrain information, but they are only practical when a single UAV serves a single user.
In this paper, we put forward two methods to avoid building blockage in a multi-user system by collecting prior terrain information and using real-time search.
We proposed four algorithms related to the combinations of the above methods and their performances are evaluated and compared in different scenarios.
By adjusting the height of the UAV based on terrain information collected before networking, the performance is significantly enhanced compared to the one when no terrain information is available.
The algorithm based on real-time search further improves the coverage performance by avoiding the shadow of buildings. During the execution of the real-time search algorithm, the search distance is reduced using the collected terrain information.

\end{abstract}


\section{Introduction}
\IEEEPARstart{A}{chieving} ubiquitous coverage for outdoor regions is one of the critical issues in the fifth generation (5G) and beyond 5G communication networks \cite{pan2023computer,shang2022UAV,lou2023haps}.
Outdoor coverage includes providing connectivity for outdoor residents, firefighters during a wildfire, rescue teams in emergency situations, and soldiers during military operations. In particular, non-terrestrial networks (NTN) platforms, comprising satellite constellations, high altitude platform stations (HAPS), and low altitude platform stations (LAPS), play an indispensable role in achieving seamless outdoor coverage \cite{lou2023haps,wang2022ultra}. Compared to conventional ground-based infrastructure, NTN platforms offer swifter construction and deployment. \cite{huang2023system,wang2023reliability}. However, outdoor coverage is challenging due to the high mobility of the crowds, the dominance of non-line-of-sight (NLoS) propagation paths \cite{imran2019seamless}, and the diverse terrain configurations \cite{du2020directional}.
The self-adaptive and self-evolving micro-networks that are expected in the sixth generation (6G) vision will provide coverage for outdoor users \cite{wang2022stochastic,letaief2019roadmap,belmekki2020unleashing,wang2023resident}. Owing to their mobility, unmanned aerial vehicle (UAV) networks are well-suited for such a highly flexible network architecture \cite{yaacoub2020key}. Furthermore, UAVs' lower operating altitude grants the benefit of reduced communication latency, which is vital for applications requiring real-time data transmission.

\par
For an energy-limited network, the power received by users will directly determine the communication quality of service \cite{wang2022evaluating,elzanaty2021towards,huang2023system}.
Moreover, the common metrics used to assess aerial networks, such as coverage probability, achievable data rate, and energy efficiency, are directly related to the users' received power \cite{wang2022ultra,lou2023coverage,khuwaja2018survey,zaid2023evtol}.
In scenarios where the UAV provides coverage for outdoor users, the user's received power can be enhanced by shortening the distance between the UAV and the user and by avoiding building blockage \cite{lou2021green,zaid2023aerial}.
Most of the research related to UAV networking attaches importance to the terrain influence such as the blockage of buildings \cite{wang2023terrain}.
Although terrain impact on performance is a major focus in UAV networks, most studies use basic models to describe the impact of building blockage on the user's performance.
In the following, we present three challenges that impede the usage of elaborated and sophisticated models.

\par
{\color{black} The first challenge is that lowering the UAV altitude reduces the distance between the UAV and the user but increases the probability of NLoS propagation. A widely applied stochastic channel model \cite{al2014optimal} provides an acceptable solution to address this challenge. The second challenge is related to the irregularity of the terrains that can significantly impact the air-to-ground (ATG) link quality. In this case, a slight movement of the UAV may lead to a significant change in channel gain \cite{gesbert2022uav}. The real-time search approach, as a new approach emerging in recent years, is leveraged to solve the fine-grained UAV deployment problem, which addresses the second challenge\cite{chen2017optimal,nawaz2021uav}. The third challenge pertains to simultaneously managing multiple users' quality of service  \cite{dang2020should}. When the UAV attempts to establish a direct path with one user, the original direct path with another user might be blocked. Consequently, when the number of users being served increases, the difficulty of deployment rises dramatically. While previous research has made significant progress in addressing the first two challenges individually, the unique challenge of building blockage in multi-user systems has remained unexplored, and strategies to address all three challenges simultaneously have yet to be provided. In order to tackle these three challenges simultaneously, we have introduced novel algorithms that leverage both prior terrain information and real-time search. The fusion of these elements represents a pioneering approach, setting our work apart from existing research. Importantly, our approach also incorporates coverage probability estimation and user classification, further underscoring its innovative nature.}

\subsection{Related Works}
Most of the existing works are based on a stochastic channel model to estimate the ATG wireless channel model \cite{feng2006wlcp2,al2014modeling,matracia2021coverage}.
The stochastic channel model answers the following question: \textit{how the relative positions between UAVs and the served users affect the probability of establishing line-of-sight (LoS) links between them?}\cite{al2014optimal,sun2023joint}.
In \cite{feng2006wlcp2}, the authors divide ATG propagation modes into two main types, i.e., LoS and NLoS, while their probability distribution is modeled as a function of the elevation angle from user to UAV, however, shadow fading is not considered.
Subsequently, the authors \cite{al2014modeling} provide a general ATG path loss prediction model based on the environment statistical parameters related to the receiver's location.
In addition, the authors \cite{al2014optimal} propose a simple modified sigmoid function with two environment parameters to approximate the probability of LoS. Although the statistical model of the ATG channel can provide an estimation for the expected service level \cite{alzenad20173}, it cannot exploit the terrain information for UAV positioning.

Compared to UAV deployment based on stochastic channel models, real-time search methods are more granular and utilize terrain information more efficiently \cite{chen2017optimal,nawaz2021uav}.
In \cite{he2018UAV}, the authors maximize the coverage area of UAVs while users are located on uneven terrain, i.e., the altitudes of users are different, but the environment is still coarse-grained.
More recently, some literature combine actual and fine-grained terrain with a UAV-assisted system in which a UAV acts as a relay for a single user and base station (BS) to extend the coverage of BS and boost the capacity of a shadowed user \cite{chen2020map,chen2021uav}.
The authors \cite{chen2020map} leverage local terrain information to optimize the placement and trajectory of the UAV.
In \cite{chen2021uav}, the authors seek the optimal UAV placement and demonstrate the superior performance of their proposed terrain-based approach in a real-world urban topology.

\par
In summary, stochastic channel model-based methods are generally inaccurate since they use the basic characteristics of the terrain.
Although the real-time search method is more effective in using terrain information, methods in existing literature only apply to the scenario where one UAV serves one user.
Fortunately, these stochastic channel model-based methods and real-time search methods are used before and during networking, respectively.
As a result, we are able to combine the benefits of the two approaches, enhance the accuracy of the former, and expand the use of the latter to accommodate numerous users.

\subsection{Contribution}
{\color{black}
This paper designs a terrain-based UAV deployment system, divided into four cases according to the availability of $(I_1)$ prior terrain information and $(I_2)$ real-time search. $(I_1)$ requires UAV to have sufficient resources for terrain data collection before providing coverage to users, while $(I_2)$ demands UAV to have sufficient electric power and a fast speed compared to the users for real-time search. The contributions related to each case are as follows:
\begin{itemize}
    \item Both $(I_1)$ and $(I_2)$ are not available: An improvement of the weighted $K$-nearest neighbor (WKNN) algorithm, called \textbf{barycenter-inspired algorithm (BIA)}, is proposed. The BIA considers the coverage characteristics of the UAV network in weight (mass density function) design.
    \item $(I_1)$ is available, and $(I_2)$ is not available: We extract from the prior terrain information as two terrain-related feature parameters using ridge regression. The \textbf{stochastic channel-based positioning algorithm (SCPA)} is proposed based on the parameters for UAV position selection. 
    \item $(I_1)$ is not available, and $(I_2)$ is available: We propose the \textbf{multi-user-based real-time search algorithm (MRSA)}, where the UAV can obtain terrain information and determine the next position while providing network services. The MRSA applies to the scenario where one UAV serves multiple users with a linear search length. In particular, the UAV deployment location obtained by MRSA is proved to be conditionally optimal when a UAV serves two users.
    \item Both $(I_1)$ and $(I_2)$ are available: With prior terrain information, we propose a resident classification method to narrow the range of users that UAV needs to serve. Based on this method, we improve MRSA and further propose the \textbf{hybrid deployment algorithm (HDA)}. Compared to MRSA, HDA requires a shorter search distance and can find the UAV deployment location with better coverage performance. 
\end{itemize}
Please note that MRSA and HDA are algorithms employed for the first time in multi-user UAV systems. Additionally, what distinguishes them from existing real-time search algorithms is the integration of probability estimation and user classification.
}

\begin{figure*}[t]
	\centering
	\includegraphics[width=0.9\linewidth]{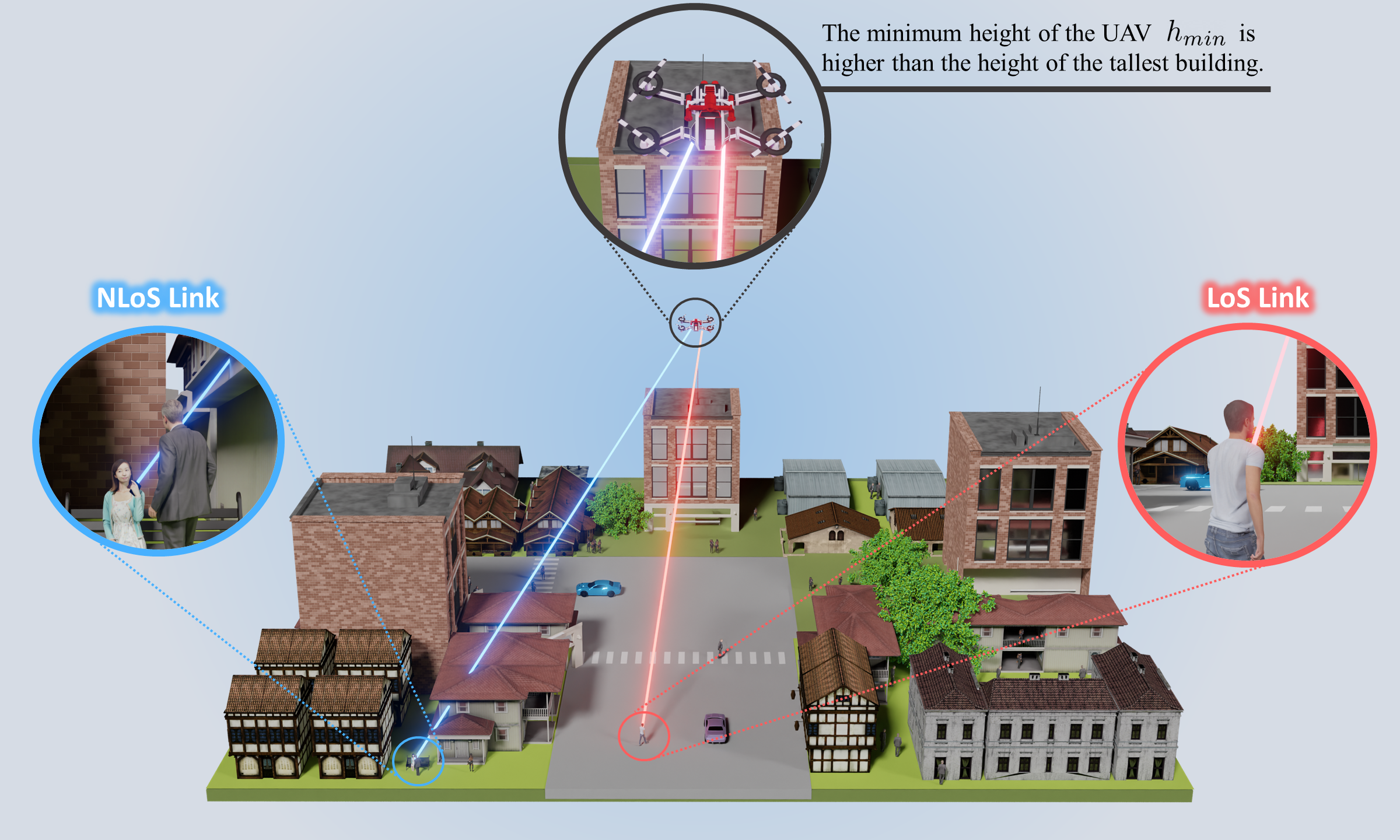}
	\caption{Illustration of the system model configurations.}
	\label{sysmodel}
	\vspace{-0.4cm}
\end{figure*}

\section{System Model}\label{section2}
In this section, we provide a scenario description, channel model, and optimization objectives related to the study. 

\subsection{Scenario Description}
We consider a suburban region containing several buildings, one UAV, and $K$ outdoor ground users. Users are required to upload their locations when requesting network access services. The UAV dynamically adjusts its position based on the users' locations. 
To avoid collisions with buildings, the minimum height of the UAV, denoted by $h_{\min}$, is higher than the tallest building as depicted in Fig~\ref{sysmodel}. The above assumptions are common and reasonable in UAV networking \cite{chen20213d}. However, the specific details of the terrain are unknown to the UAV since most of the available databases do not provide detailed terrain information, such as the height and shape of buildings as well as other signal barriers such as large vehicles. Finally, we assume that there are no other public access points (APs) in the area, and outdoor users can only access the network via the UAV. Then, the UAV establishes a dependable LoS connection with the core network to meet the backhaul requirements, owing to the higher altitudes at which both the UAV and base station are typically deployed compared to the surrounding buildings.

\subsection{Channel Model}
The signal received by the user is considered in a LoS when no building is blocking the signal between the user and the UAV. Otherwise, the signal is considered in a NLoS.
The UAV-user channel undergoes large-scale and small-scale fading. Therefore, the received power $S_{Q}$, $Q=\left\{\rm{LoS},\rm{NLoS}\right\}$ is expressed as
\begin{equation}\label{S_Q}
    S_{Q} \left(r\right) =
    \eta_{Q} \, \zeta  \, G_{Q} r^{-\alpha_{Q}}, 
\end{equation}
where $\zeta $ is the transmit power, $\eta_{Q}$ is the mean additional losses \cite{UAV_SG}, and $r$ is the Euclidean distance between the user and the UAV. The large-scale fading is denoted by $\eta_{Q} \, r^{-\alpha_{Q}}$, where $\alpha_{Q}$ is the path-loss exponent. Considering the significant multipath effect of buildings, the small-scale fading $G_{Q}$ is modeled by a Nakagami-$m$ fading channel with a shape parameter and scale parameter $(m_{Q}, \frac{1}{m_{Q}})$. The probability density functions (PDF) of the power gains $G_{Q}$ is given by \cite{nakagami}
\begin{equation}
    {f_{G_{Q}}}\left(g\right) = \frac{{m_{Q}}^{{m_{Q}}}{g^{{m_{Q}} - 1}}}{{\Gamma \left( {m_{Q}} \right)}}{e^{ - {m_{Q}} \cdot g}},
\end{equation}
where $\Gamma \left( {{m_{Q}}} \right) = \int_0^\infty  {{x^{{m_{Q}} - 1}}{e^{-x}}dx}$ is the Gamma function.
\par

Therefore, the overall coverage probability, which is the probability that the received  signal-to-noise ratio (SNR) is greater than a predefined threshold $\gamma$, can be derived by 
\begin{equation}\label{PC_definition}
    P_C \left(h;r_1,r_2,...,r_K\right) = \frac{1}{K} \sum_{k=1}^K \mathbb{P} \left[ \frac{S_{Q,k}\left(r_k\right)}{\sigma^2}>\gamma \right],
\end{equation}
where $h$ is the height of the UAV, $r_k$ is the Euclidean distance between the $k^{th}$ user and the UAV, $\sigma^2$ is the environmental noise received by the user, and $Q$ is either LoS or NLoS for a given user $k$. 

\subsection{Optimization Objectives}
Our goal is to determine the UAV position that maximizes the overall coverage probability $P_C$, which we define as \textit{the optimal position}.
\begin{definition}[Terrain-based Optimal Position]
The position that has the maximum overall coverage probability within the three-dimensional region accessible to the UAV is called a terrain-based optimal position (simply denoted as optimal).
\end{definition}

Considering the above channel model, the optimal position can be determined using the following two approaches: reducing the distance $r_k$, and establishing LoS link rather than NLoS link with user $k$. Then, our optimization problem can be formulated as
\begin{subequations} 
	\begin{alignat}{2}
		&\underset{ \boldsymbol{h; r_k, 1\leq k\leq K}}{\text{maximize}}      &\qquad &  \frac{1}{K} \sum_{k=1}^K \mathbb{P} \left[ \frac{S_{Q,k}\left(r_k\right)}{\sigma^2}>\gamma \right], \\
		&\textrm{subject to:}    &      & h>h_{\min}, \label{Constraint1}
	\end{alignat}
\end{subequations}
where the objective function is the overall coverage probability that can be seen as a weighted sum of users' coverage probability such that the weight of each user is the same. The height of the UAV $h$, and the Euclidean distance between the UAV and each user $r_k, 1 \leq k \leq K$, are treated as optimization variables. The only constraint  (\ref{Constraint1}) is enforced to prevent collisions between the UAV and buildings. Finally, it is noted that the position with the highest instant coverage probability or received signal power is not necessarily the optimal position on the trajectory because of the small-scale fading fluctuations. Therefore, the UAV can only utilize the instant received signal power to check the blockage in real-time search.

\section{The Acquisition and Processing of Terrain Information}\label{section3}

\subsection{LoS Probability Estimation}
\label{obtain blockage}
In this subsection, we first introduce the LoS probability expression with undetermined parameters, then proceed to describe how these parameters can be estimated according to the data collected by UAV. Finally, we discuss the LoS probability in more detail.
\par
We consider that the UAV can obtain the characteristics of the LoS probability by collecting a small amount of data in advance, but such data is not enough for reconstructing the terrain. The authors in \cite{al2014optimal,alzenad20173} express the probability of establishing a LoS link as a modified sigmoid function
\begin{equation}
\label{PLoSk}
    P_{\rm{LoS}}\left( h,r \right) = \frac{1}{{1 + a \exp \left( { - b \left( {\frac{{180}}{\pi }{{\tan }^{ - 1}}\left( {\frac{{{h}}}{ \sqrt{r^2-h^2} }} \right){-a}} \right)} \right)}}.
\end{equation}

Based on the above function, the relationship between $P_{\rm{LoS}}\left( h,r \right)$ and $h,r$ can be determined by two terrain-related parameters, $a$ and $b$. The values of $a$ and $b$ are related to the density, height, and floor area of the buildings. Since the establishment of LoS link and NLoS link are mutually exclusive events, the corresponding two probabilities are complementary, that is, $P_{{\rm{NLoS}}}\left( h,r \right) = 1 - P_{{\rm{LoS}}}\left( h,r \right).$

Next, we explain how to collect the data set and estimate parameters $a$ and $b$ based on the data set. In order to facilitate data collection, we define the elevation angle $\theta$ as
\begin{equation}\label{theta}
    \theta = \frac{{180}}{\pi }{{\tan }^{ - 1}}\left( {\frac{{{h}}}{ \sqrt{r^2-h^2} }} \right).
\end{equation}

For the data collection of a fixed $\theta=\theta_i$, the height of the UAV $h$ is adjusted to satisfy the relationship between $h$, $\theta_i$ and the horizontal distance from the UAV to a randomly chosen user $\sqrt{r^2-h^2}$ in (\ref{theta}). 
In order to ensure fairness during the data collection process, the UAV modifies its position and associates various users while maintaining a fixed $\theta_ i$. 
Then, the ratio between the number of samples of LoS link established and the total number of samples is obtained, denoted as $t_i$. Finally, the above steps are repeated by changing $\theta$ to obtain a data set containing $N$ tuples: $\mathcal{T} = \{ (t_1,\theta_1), (t_2,\theta_2), \dots , (t_N,\theta_N) \}$.
\par
Based on the data set $\mathcal{T}$, we choose the mean square error (MSE) function as the cost function. Regularization is added in order to enhance the generalization ability and to reduce the influence of the edge effect (non-uniform sampling in edge regions due to limited map size). Therefore, the parameter estimation problem is described by the following nonlinear ridge regression problem
\begin{equation}\label{optimization}
    \left( a^* , b^* \right) = \underset{ {a, b}}{\arg \min} \sum_{i=1}^N \left( t_i - \frac{1}{{1 + a \exp \left( { - b \left( {\theta_i - a} \right)} \right)}} \right)^2 
    + \lambda_1 \left(a-\widehat{a}\right)^2 + \lambda_2 \left(b-\widehat{b}\right)^2,
\end{equation}
where $\widehat{a}$ and $\widehat{b}$ are empirical parameters that provided in \cite{bor2016efficient}, $\lambda_1$ and $\lambda_2$ are regularization coefficients. Since $\widehat{a}$ is much larger than $\widehat{b}$, $\lambda_2$ should also be much larger than $\lambda_1$ to ensure that both arguments have similar learning rates.

\par
Finally, we add two relevant comments to the LoS probability estimation problem. Firstly, the modified sigmoid function in (\ref{PLoSk}) may not be the only choice, and good estimates can be made by improving other commonly used machine learning functions. We improve the Tanh and ReLu functions and also use only two parameters to estimate the LoS probability. 
Secondly, we apply the trust region reflective algorithm to solve the optimization problem in (\ref{optimization}), and the initial point is set as $(\widehat{a}, \widehat{b})$.

\subsection{Coverage Probability and Resident Classification}\label{classification}
In this subsection, we first give the motivation for classification. Then, a closed-form expression for the coverage probability is derived. Finally, we show how the users are classified based on the coverage probability. 
\par
The authors in \cite{zou2005distributed} introduce a probabilistic node sensing model and divide the receivers into three classes according to their distances from the transmitter: definitely covered ($C_1$), probably covered ($C_2$), and cannot be covered ($C_3$). Through the classification, the UAV should focus on $C_2$ users to improve the overall coverage, 
because it does not need to focus on $C_1$ (close) users and the $C_3$ (distance) users, since the former are already covered, the latter cannot be covered.

\par
The second motivation for user classification is that the coverage probability of $C_2$ users decreases almost linearly with the increase of the Euclidean distance between the UAV and the user. Using this important property, the expression of overall coverage probability can be simplified in some cases. Based on (\ref{S_Q}) and (\ref{PLoSk}), the coverage probability is given in the following theorem.

\begin{lemma}\label{P_C}
Given that the height of UAV is $h$ and distance between the $k^{th}$ user and the UAV is $r_k$, the coverage probability of the $k^{th}$ user is given by
\begin{equation}
    P_{C,k} \left( h , r_k \right) = P_{\rm{LoS}} \left( h , r_k \right) P_{C,k}^{\rm{LoS}}\left( r_k \right) + P_{\rm{NLoS}} P_{C,k}^{\rm{NLoS}}\left( r_k \right) ,
\end{equation}
where $P_{\rm{LoS}}\left( h , r_k \right)$ is defined in (\ref{PLoSk}), and $P_{{\rm{NLoS}}}\left( h,r_k \right) = 1 - P_{{\rm{LoS}}}\left( h,r_k \right).$ $P_{C,k}^Q\left( h , r_k \right)$ is the coverage probability under the condition that $Q$ link is established, where $Q=\left\{\rm{LoS},\rm{NLoS}\right\}$. The expression of $P_{C,k}^Q\left( h , r_k \right)$ is given by
\begin{equation}\label{conditional coverage}
\begin{split}
    P_{C,k}^Q\left( r_k \right) = \exp\left( - \mu_Q \left( r_k \right) \right) \sum_{n=0}^{m_Q-1} \frac{\mu_Q \left( r_k \right)^n}{n!},
\end{split}
\end{equation}
where $m_Q$ is shape parameter of Nakagami-$m$ fading and $\mu_Q \left( r_k \right)$ is given by
\begin{equation}\label{mu}
    \mu_Q \left( r_k \right) = \eta_{Q}^{-1} \zeta ^{-1} \gamma \, r_k^{\alpha_Q} \sigma^2.
\end{equation}
\begin{proof}
See Appendix \ref{app:P_C}.
\end{proof}
\end{lemma}

The conditional coverage probability expression of in (\ref{conditional coverage}) contains the exponential term \\ $\exp\left( - \mu_Q \left( r_k \right) \right)$, therefore, when $\mu_Q \left( r_k \right) \gg 1$, the conditional coverage probability converges rapidly to 0; whereas when $0 < \mu_Q \left( r_k \right) \ll 1$, the conditional coverage probability converges rapidely to 1. This clarifies the reasoning behind the above classification. Only a small range of $r_k$ makes $ \mu_Q \left( r_k \right) \approx 1$ and brings a significant change in the conditional coverage probability. 
\par
Note that the coverage probability is related to the LoS probability obtained in subsection \ref{obtain blockage}, while the conditional coverage probability is not. Accordingly, we denote by $\varepsilon$ the degree of classifications, and we provide the following two types of classifications:
\begin{itemize}
    \item Non-terrain-based classification: When $P_{C,k}^{\rm{NLoS}}\left( r_k \right) > 1 - \varepsilon$, user $k$ belongs to $C_1$; when $P_{C,k}^{\rm{LoS}}\left( r_k \right) < \varepsilon$, user $k$ belongs to  class $C_3$; otherwise user $k$ is classified into the class $C_2$.
    \item Terrain-based classification: When $P_{C,k}\left( r_k \right) > 1 - \varepsilon$, user $k$ belongs to $C_1$; when $P_{C,k}\left( r_k \right) < \varepsilon$, user $k$ belongs to  class $C_3$; otherwise when $\varepsilon \leq P_{C,k}\left( r_k \right) \leq 1 - \varepsilon$, user $k$ is classified into the class $C_2$.
\end{itemize}
\par
  

\subsection{Real-Time Search in Simple Scenarios}\label{real-time}
In this subsection, we will start with the real-time terrain-based search for two simple scenarios. The conclusion obtained from the following analysis of simple scenarios can inspire the design of the MRSA and HDA.

\begin{itemize}
    \item Scenario~\uppercase\expandafter{\romannumeral1}: There are no buildings, and the height of the UAV's optimal position is, therefore, the lowest altitude $h_{\min}$. This scenario is analyzed in order to find a feasible optimization objective.
    
    \item Scenario~\uppercase\expandafter{\romannumeral2}: The UAV provides coverage for only two users. Based on the analysis of scenario~\uppercase\expandafter{\romannumeral2}, a practical search strategy is proposed.
    
\end{itemize}

For the above two scenarios, the UAV searches along a trajectory specified by a strategy $\mathscr{L}$ and chooses a point on the trajectory as the final positioning result. For the above search methods (MRSA and HDA), the two characteristics that we are most interested in are optimality, defined in section~\ref{section2}, and linearity. 

\begin{definition}[Linearity of the Trajectory]
Linearity is a function of (i) the distance among the users, and (ii) the UAV height. For example, the distance among the users increases, (the upper bound of) the length of the trajectory of $\mathscr{L}$ increases linearly, in that case, the strategy $\mathscr{L}$ is called a linear strategy.
\end{definition}

Furthermore, from both latency and energy perspectives, the length of trajectory is highly limited. Therefore, the linear search length of the UAV is necessary. Compared to the flight duration of the UAV, the computational delay is significantly shorter, so we place a greater emphasis on the analysis of UAV search length. Combining linearity and optimality, the following definition is given.

\begin{definition}[Linear Optimal Strategy]
If a linear strategy $\mathscr{L}$ can always pass the optimal position, then we call $\mathscr{L}$ a linear optimal strategy.
\end{definition}

A squared optimal strategy, which involves searching the whole plane to find the optimal solution, exists in scenario \uppercase\expandafter{\romannumeral1}. However, whether a linear optimal strategy, which aims to find the optimal solution by searching the space incrementally with a linear growth pattern, exists in scenario \uppercase\expandafter{\romannumeral1} is discussed in the following proposition.

\begin{proposition}\label{proposition1}
In scenario~\uppercase\expandafter{\romannumeral1}, the relationship between the existence of linear optimal trajectory $\mathscr{L}$ and the number of users is as follows:
\begin{itemize}
    \item When the UAV provides coverage to one or two users, a linear optimal trajectory always exists.
    \item When the UAV provides coverage to three users, an approximate linear optimal trajectory can be found by approximating the problem of searching the optimal position to the problem of finding the Fermat-Weber point \cite{carmi2005fermat}.
    \item When the UAV provides coverage to more than four users, the linear optimal trajectory does not exist. 
\end{itemize}
\begin{proof}
See Appendix~\ref{app:proposition1}.
\end{proof}
\end{proposition}

Although the above analysis can provide inspiration for the real-time terrain-based search algorithm, it is challenging to provide effective search trajectories for large numbers of users. Therefore, we define the average SNR as
\begin{equation}
    \overline{\rm{SNR}}_Q(r) = \mathbb{E}_G \left[ \frac{S_Q(r)}{\sigma^2} \right] = \frac{\eta_Q \zeta  r^{\alpha_Q}}{\sigma^2},
\end{equation}
where $Q=\{ \rm{LoS} , \rm{NLoS} \}$. Note that optimizing other performance metrics, such as average data rate and energy efficiency \cite{zhang2016energy}, ultimately boils down to optimizing the average SNR. While our focus lies on optimizing overall coverage probability, it indirectly impacts and improves these associated metrics. The following definition is given to simplify the optimization objective by seeking the linear suboptimal trajectory instead of the linear optimal one.

\begin{definition}[Linear $\gamma$-suboptimal Strategy]
A strategy $\mathscr{L}$ is said to be linear $\gamma$-suboptimal (simply denoted as $\gamma$-suboptimal) if (\romannumeral1) there is one and only one $\gamma$-suboptimal position that satisfies the condition that the average SNR from this position to any user in $C_2$ is greater than or equal to $\gamma$, (\romannumeral2) $\mathscr{L}$ can find the $\gamma$-suboptimal position, and (\romannumeral3) $\mathscr{L}$ is a linear strategy.
\end{definition}

\begin{proposition}\label{proposition2}
In scenario \uppercase\expandafter{\romannumeral1}, there always exists a $\gamma$-suboptimal strategy $\mathscr{L}$. 
\begin{proof}
See Appendix~\ref{app:proposition2}.
\end{proof}
\end{proposition}

Next, we will propose a $\gamma$-suboptimal strategy in scenario~\uppercase\expandafter{\romannumeral2}. Since the UAV provides coverage for only two
users, we establish a cylindrical coordinate system. Take the midpoint of the two users as the origin, the line passing through the two users as $z$-axis, and the direction perpendicular to the ground as the direction of $\theta=\frac{\pi}{2}$. Therefore, the users' cylindrical coordinates are $(\rho,\theta,z)=(0, 0, \pm\frac{d}{2})$, where $d$ is the distance between two users. 

\begin{figure*}[t]
	\centering
	\includegraphics[width=0.99\linewidth]{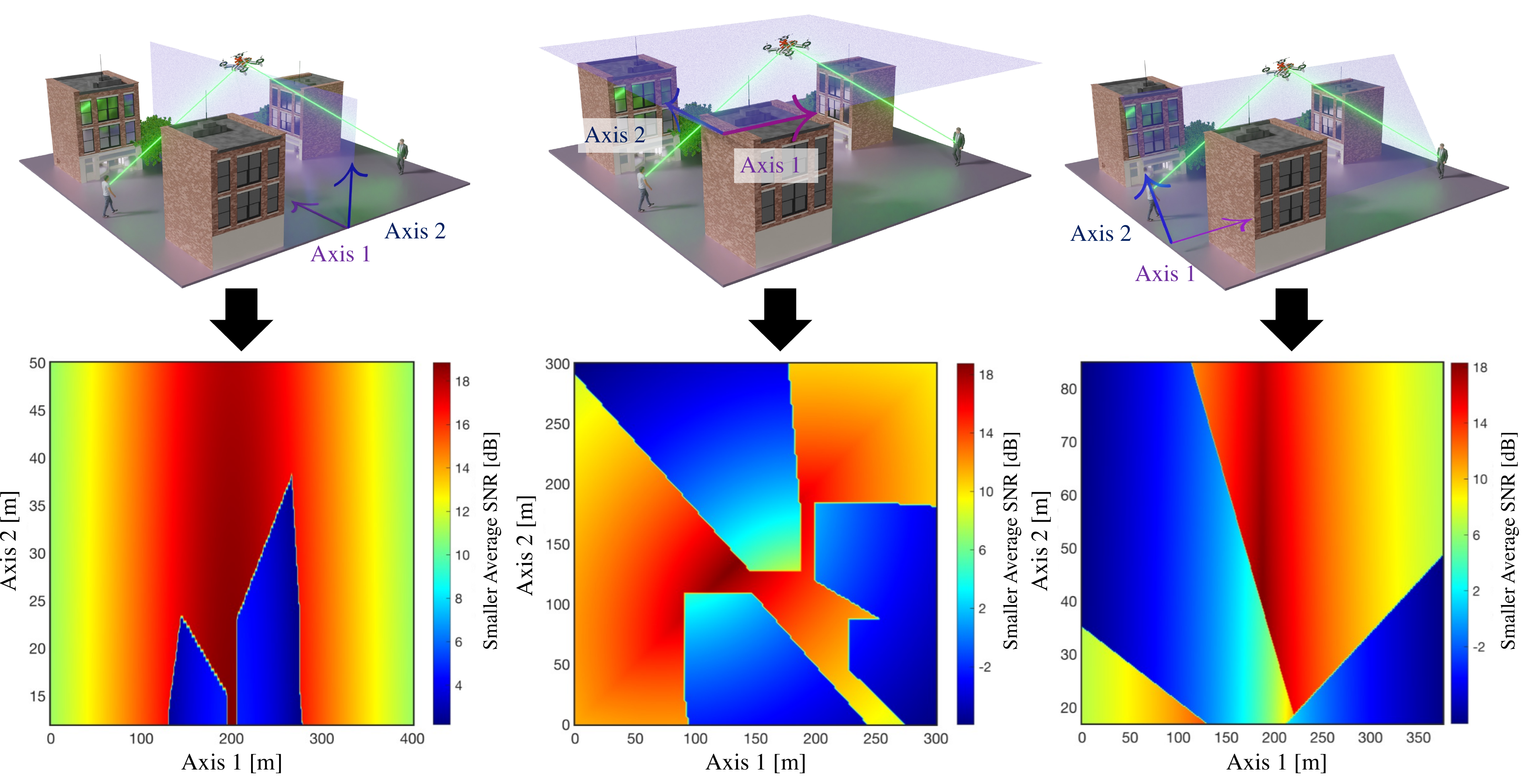}
	\caption{Heat maps of smaller average SNR between two users from different perspectives.}
	\label{fig:big}
	\vspace{-0.4cm}
\end{figure*}

Fig.~\ref{fig:big} shows heat maps of smaller average SNR between two users from different perspectives. There are some distinct boundaries in the figure, indicating that one user switches from not being blocked to being blocked, while the other user is never blocked by the building. From Fig.~\ref{fig:big}, we know that only in the $z=0$ plane, the LoS region is continuous because any point below a point in the NLoS region also belongs to the NLoS region. Given the special property of $z=0$ plane, the following strategy is proposed: (\romannumeral1) search on the $z=0$ plane, (\romannumeral2) decrease $\rho$ until at least one user's link to the UAV is blocked by the building and (\romannumeral3) change the value of $\theta$ 
until both links (between the UAV and the two users) are in LoS.

 \begin{algorithm}[!ht] 
	\caption{Real-Time Search Algorithm for Two Users}
	\label{alg1} 
	\begin{algorithmic}[1]
		
		\STATE \textbf{Input}: The starting coordinates of the UAV $(\rho_u, \theta_u, 0)$, search granularity $\delta$.
		
		\STATE \textbf{Initialize}: Boolean variable $\mathcal{B} \leftarrow 0$ if both of the users establish LoS links with the UAV, otherwise $\mathcal{B} \leftarrow 1$. Let $z_u \leftarrow 0$.

		\WHILE{$\mathcal{B} = 1$}
		\STATE $\rho_u \leftarrow \rho_u + \delta$, and renew the value of $\mathcal{B}$.
		\ENDWHILE
		
		\WHILE{$\rho_u \cos\theta_u > h_{\min}$}
		
		\IF {$\mathcal{B} = 1$}
		\STATE $\rho^* \leftarrow \rho_u$, $\theta^* \leftarrow \theta_u$, $\rho_u \leftarrow \rho_u - \delta$.
		
		\ELSE
		\STATE $\theta_u \leftarrow \theta_u - 2 \arcsin \frac{\delta}{2 \rho_u}$.
		\ENDIF
		
		\STATE Renew the value of $\mathcal{B}$.
		\ENDWHILE
		
		\STATE $\rho_u \leftarrow \rho^*$, $\theta_u \leftarrow \theta^*$.
		
		\WHILE{$\rho_u \cos\theta_u > h_{\min}$}
		
		\IF {$\mathcal{B} = 1$}
		\STATE $\rho^* \leftarrow \rho_u$, $\theta^* \leftarrow \theta_u$, $\rho_u \leftarrow \rho_u - \delta$.
		
		\ELSE
		\STATE $\theta_u \leftarrow \theta_u + 2 \arcsin \frac{\delta}{2 \rho_u}$.
		\ENDIF
		
		\STATE Renew the value of $\mathcal{B}$.
		\ENDWHILE
		
		\IF {$\overline{\rm{SNR}}_{\rm{NLoS}}\left(\sqrt{h_{\min}^2+\frac{d^2}{4}}\right) > \overline{\rm{SNR}}_{\rm{LoS}}\left(\sqrt{{\rho^*}^2+\frac{d^2}{4}} \right)$}
		
		\STATE $\gamma \leftarrow \overline{\rm{SNR}}_{\rm{NLoS}}\left(\sqrt{h_{\min}^2+\frac{d^2}{4}}\right)$, $\rho^* \leftarrow h_{\min}$, $\theta^* \leftarrow 0$.
		
		\ELSE 
		
		\STATE $\gamma \leftarrow \overline{\rm{SNR}}_{\rm{LoS}}\left(\sqrt{{\rho^*}^2+\frac{d^2}{4}} \right)$.
		
		\ENDIF
		
		\STATE \textbf{Output}: Position ($\rho^*, \theta^*, 0$) and $\gamma$.
	\end{algorithmic}
\end{algorithm}	

Steps (3) - (6) of Algorithm~\ref{alg1} ensure that the starting position is in the LoS region for both users. Steps (7) - (15) search the left branch, and (16) - (25) search the right branch. {\color{black} The steps (6) and (15) respectively verify whether the UAV is below $h_{\min}$ in the left branch search and the right branch search.}

\par
{\color{black} As for the final position of the UAV search trajectory, there are two potential $\gamma$-suboptimal positions in the LoS and NLoS regions. Steps (23) - (27) compare these two positions. The potential $\gamma$-suboptimal position in the NLoS region is $(h_{\min},0,0)$ (given that the position $(h_{\min},0,0)$ is not located in the common LoS region of both users). In the LoS region, the potential $\gamma$-suboptimal position is the last position in the trajectory that the UAV stays in the common LoS region of both users.}

\par
Even though the average SNR is not available to the UAV, it can infer whether the users are blocked or not. {\color{black} Furthermore, for high-rise building scenarios \cite{al2014modeling}, there may be situations where the UAV's altitude becomes too high. In such cases, we can set an upper limit for $\rho_u$. During the execution of step (4), if this limit is exceeded, the algorithm terminates and outputs $(\frac{d_{\min}}{\cos \theta_u}, \theta_u, 0)$ as the deployment position for the UAV.}

\begin{proposition}\label{proposition3}
In scenario \uppercase\expandafter{\romannumeral2}, we have the following statements for Algorithm~\ref{alg1} and its output $(\rho^*, \theta^*, 0)$:
\begin{itemize}
    \item When $2 \rho^* \leq d$, the strategy proposed by Algorithm~\ref{alg1} is $\gamma$-suboptimal. In other words, $(\rho^*, \theta^*, 0)$ is the only position where the average SNRs of both users are greater than or equal to $\gamma$.
    
    \item When $2 \rho^* > d$, the strategy proposed by Algorithm~\ref{alg1} is not $\gamma$-suboptimal and $\gamma < \overline{\rm{SNR}}_{\rm{LoS}}\left( \frac{2 \rho^* d}{\sqrt{\left(2\rho^*\right)^2 + d^2}} \right)$. 
\end{itemize}
\begin{proof}
See Appendix~\ref{app:proposition3}.
\end{proof}
\end{proposition}

\section{Typical Scenarios Algorithms Design}\label{section4}
In this section, four typical scenarios and the corresponding four positioning algorithms are shown in Table \ref{table1}. Prior information available indicates that the UAV is allowed to collect data and determine the prior probability of establishing LoS link in advance. Since the users move slower than UAV, the latter has a time span to search for the best location based on the terrain. Specifically, at each time interval, based on users' locations, the UAV adjusts its own location by recalculating its position using the corresponding algorithm in different scenarios. At last, the granularity $\delta$ is defined for the positioning accuracy.

\begin{table}[h]
\centering
\caption{Positioning algorithms for four typical scenarios.} 
\label{table1}
\begin{tabular}{|c|c|c|c|}
\hline
Algorithm &
  \begin{tabular}[c]{@{}c@{}}Prior information required\end{tabular} &
  \begin{tabular}[c]{@{}c@{}}Terrain information  utilized\end{tabular} &
  \begin{tabular}[c]{@{}c@{}}Real-time  search required\end{tabular} \\ \hline
BIA   & No  & No  & No  \\ \hline
\begin{tabular}[c]{@{}c@{}} SCPA \end{tabular} & Yes & Yes & No  \\ \hline
MRSA    & No  & Yes & Yes \\ \hline
HDA      & Yes & Yes & Yes \\ \hline
\end{tabular}
\end{table}

\subsection{Barycenter-Inspired Algorithm}
The Algorithm~\ref{alg2}, that is, the BIA is designed for the scenario when prior information is not available. It can be regarded as a variant of the clustering algorithm.
\par
We first establish a Cartesian coordinate system with the origin on the ground and the $k^{th}$ user's location can be expressed as $\left( x_k^r,y_k^r,0 \right)$. After that, we give different weights $\Lambda \left( r \right)$ to users who are at different distances from the UAV, and the weight $\Lambda \left( r \right)$ is called the mass density. In each iteration, the UAV takes the user's barycenter as its horizontal position. After adjusting the position of the UAV, the weight $\Lambda \left( r \right)$ also changes, which further leads to the change of barycenter. Therefore, the position of the UAV needs to be updated. According to the above deceptions, the BIA is given as follows.

\begin{algorithm}[!ht] 
	\caption{Barycenter-inspired algorithm (BIA).}
	\label{Barycenter-inspired algorithm}
	\begin{algorithmic} [1]
		
		\STATE \textbf{Input}: Locations of users $\left( x_k^r,y_k^r,0 \right)$, mass density function $\Lambda \left( r \right)$, maximum number of iterations $N$ and granularity $\delta$.
		
		\STATE \textbf{Initialize}: deploy the UAV at $\left( x_0^u,y_0^u,z_0^u \right) \leftarrow \left( \sum_{k=1}^K \frac{x_k^r}{K}, \sum_{k=1}^K \frac{y_k^r}{K}, h \right)$, initialize $n \leftarrow 1$.
		
		\REPEAT
		\FOR{$k = 1 : K$}
		\STATE $r_k \leftarrow \sqrt{\left(x_{n-1}^u - x_k^r\right)^2 + \left(y_{n-1}^u - y_k^r\right)^2 + h^2}$.
		\ENDFOR
		
		\STATE $\left( x_n^u,y_n^u,z_n^u \right) \leftarrow \left(  \frac{\sum_{k=1}^K \Lambda \left( r_k \right) x_k^r}{ \sum_{k=1}^K \Lambda \left( r_k \right) }, \frac{\sum_{k=1}^K \Lambda \left( r_k \right) y_k^r}{ \sum_{k=1}^K \Lambda \left( r_k \right) }, h \right)$.
		
		\STATE $n \leftarrow n + 1$.

		\UNTIL{ $n \geq N \ \OR \ \sqrt{\left(x_n^u - x_{n-1}^u \right)^2 + \left(y_n^u - y_{n-1}^u \right)^2} \leq \delta$.}
		
		\STATE \textbf{Output}: The final location of UAV $\left( x_n^u,y_n^u,z_n^u \right)$.
	\end{algorithmic}
 \label{alg2}
\end{algorithm}	
There are three remarks on the BIA. Firstly, since the height and distribution of buildings are unknown to the UAV, the flight height $h$ cannot be optimized by the distribution of users. Therefore, the algorithm only deploys the horizontal position of UAV. {\color{black} Secondly, BIA and Algorithm~\ref{alg1} are expressed by different coordinate systems. Since BIA is an improvement of the WKNN algorithm, it is more suitable to be presented in a Cartesian coordinate system. In order to facilitate the representation of contour lines in Proposition 3 and the proof of $\gamma$-suboptimal positions, Algorithm~\ref{alg1} utilizes a cylindrical coordinate system for modeling.} Finally, we compare several segmented mass density functions $\Lambda(r)$ in section \ref{numerical}.

\subsection{Stochastic Channel-Based Positioning Algorithm}
The SCPA is designed for the scenario when prior information (the LoS probability distribution) is available. Given the assumption that the computing power is sufficient, the SCPA is designed as an improvement of the brute force method. 
\par
The algorithm set the mean value of user coordinates as the initial point. 
Next, the algorithm searches for the position with the maximum total coverage probability $\sum_{k=1}^K P_{C,k} \left( h, r_k \right)$ around the initial point through brute force. Due to the continuity of user movement, the position of the UAV at this moment is usually close to the position of the previous moment. Therefore, the initial point can also be set to the position of the previous moment, and the search region of the brute force method does not need to be large.
\par
The similarity between the above two algorithms is that the SCPA can be regarded as the BIA with a different and inexplicit nonlinear mass density function. The difference between the two algorithms is that the SCPA optimizes the height of UAV.


\begin{figure}[t!]
	\centering
	\includegraphics[width=0.6\linewidth]{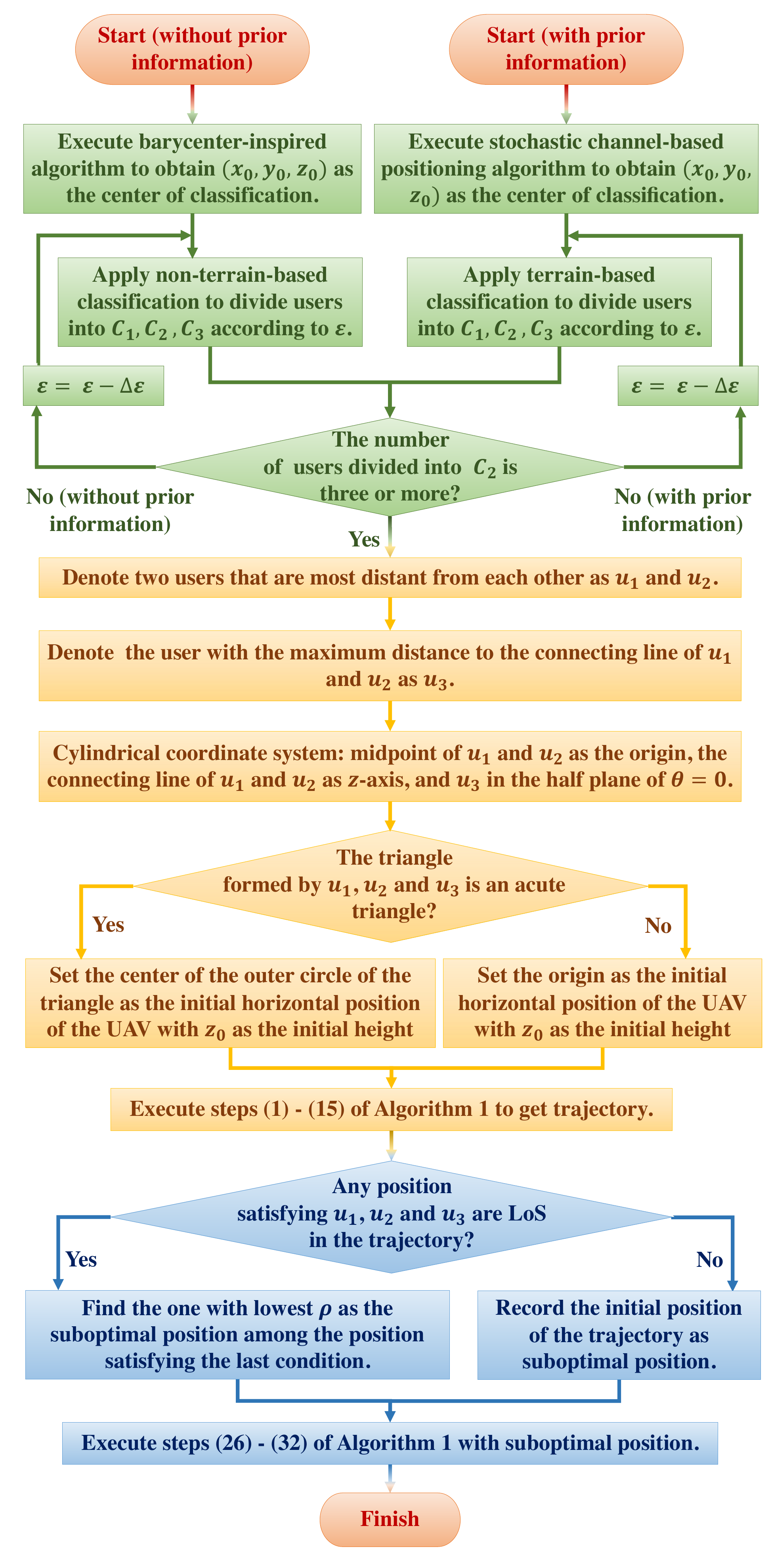}
	\caption{Flow chart of algorithms required for MRSA and HDA.}
	\label{fig:alg}
	\vspace{-0.4cm}
 \end{figure}
 
\subsection{Multi-user-based Real-time
Search Algorithm and Hybrid Deployment Algorithm}

In this subsection, we first summarize the search strategy of the simple scenarios described in subsection~\ref{real-time}, and then extend the above strategies into a general one. Since the linear optimal strategy is difficult to find even in simple scenarios, we focus on the linear $\gamma$-suboptimal strategy and provide the following deductions.

\begin{itemize}
    \item Deduction from Proposition~\ref{proposition1}: When there are less than three users in $C_2$, the Fermat-Weber point of users with the height of $h_{\min}$ can be chosen as the potential optimal position, regardless of the building blockage. 
    \item Deduction from Proposition~\ref{proposition2} and Appendix~\ref{app:proposition2}: The minimum enclosing circle can be determined by first finding the two furthest apart users in $C_2$, then repeatedly shrinking the circle and replacing the boundary users. The center of the smallest enveloping circle is taken as the possible horizontal position of the UAV, regardless of the blockage of buildings.

    \item Deduction from Proposition~\ref{proposition3}: In most cases, the trajectory proposed by Algorithm~\ref{alg1} is able to avoid the blockage and find the $\gamma$-suboptimal position when there are only two users. Even if the $\gamma$-suboptimal position cannot be found, the algorithm can gradually improve the coverage of two users during searching and provide a possible upper bound for $\gamma$.
\end{itemize}

Based on the above analysis, we propose algorithms under real-time search, which apply to a general scenario, that is, at least three users need coverage. The algorithms under real-time search contain the MRSA and HDA, and the differences between them are reflected in green steps shown in Fig.~\ref{fig:alg}. To obtain the center of classification $\left(x_0,y_0,z_0\right)$, MRSA and HDA will execute BIA and SCPA, respectively. In user classification, the MRSA and HDA will apply non-terrain-based classification and terrain-based classification, respectively. Then the second part (yellow steps in Fig.~\ref{fig:alg}) are inspired by Proposition~\ref{proposition2}, and the third part (blue steps in Fig.~\ref{fig:alg}) are inspired by Proposition~\ref{proposition3}. Finally, when only one or two users require networking, the above algorithm can be substituted by hovering directly above the users (inspired by Proposition~\ref{proposition1}) or by executing Algorithm~\ref{alg1}.

\par
{\color{black}
Finally, since the impact of $\varepsilon$ on the coverage performance of MRSA and HDA is difficult to express directly through explicit expressions, we provide the following criteria for determining the values of $\varepsilon$. 
\begin{itemize}
    \item When the steep slope portion of the curve that depicts coverage probability as a function of the UAV-user distance is not categorized as $C_2$ (examples are given in Fig.~\ref{fig7}), the value of $\varepsilon$ should be reduced. 
    \item When the search trajectory becomes lengthy, causing the UAV to be unable to perform a round of MRSA or HDA before the next update of user positions, the value of $\varepsilon$ should be increased.
    \item When a large number of users are classified as $C_2$, making it challenging for the UAV to locate the common LoS region of these users, the value of $\varepsilon$ should be increased.
\end{itemize}
}



\section{Numerical Results}\label{numerical}
In this section, we analyze the characteristics of the four algorithms and compare their performances. In each simulation, we run the code for $10^4$ rounds. In each round, we first generate the user's location based on the Poisson point process (PPP) within an area of $300~$m$~\times~300~$m, then we use one of the four algorithms to obtain the position of the UAV, and finally, we place the UAV at this position and calculate the proportion of the user under coverage. In the simulation, the buildings are modeled as prisms\footnote{The latitude and longitude of the buildings' location can be downloaded from the website www.openstreetmap.org. This paper also converts them into coordinates in the Cartesian coordinate system.}, and their heights follow a Rayleigh probability distribution with suburban's parameter \cite{al2014optimal}. Note that we have simplified the representation of buildings in simulation due to the intricacy of actual terrain construction, but in the majority of real-world scenarios, our algorithm remains effective in the presence of inclined or irregular shapes buildings.
For the sake of fairness, the algorithms are compared based on the same value of granularity $\delta=1~$m. For the sake of convenience, the simulation parameters are provided
in Table~\ref{tab_p}, unless specified otherwise.

\begin{table}[h]
\centering
\caption{System configuration.}
\label{tab_p}
\begin{tabular}{|c|c|c|}
\hline
Parameter                 & Symbol      & Value     \\ \hline \hline
Transmit power            & $\zeta $    & $30$~dBm  \\ \hline
Environmental noise power & $\sigma ^2$ & $-98$~dBm \\ \hline
Threshold for SNR         & $\gamma$    & $22$~dB   \\ \hline
Path-loss exponents of LoS and NLoS                  & $\alpha_{\rm{LoS}}$, $\alpha_{\rm{NLoS}}$ & $2$, $2.3$         \\ \hline
Nakagami-$m$ fading shape parameters of LoS and NLoS & $m _{\rm{LoS}}$, $m _{\rm{NLoS}}$          & $2$, $1$           \\ \hline
Mean additional losses of LoS and NLoS               & $\eta  _{\rm{LoS}}$, $\eta  _{\rm{NLoS}}$  & $-35$~dB, $-48$~dB \\ \hline
\end{tabular}
\end{table}

\begin{figure}[h]
\begin{minipage}[t]{0.46\linewidth}
\centering
\includegraphics[width=0.98\linewidth]{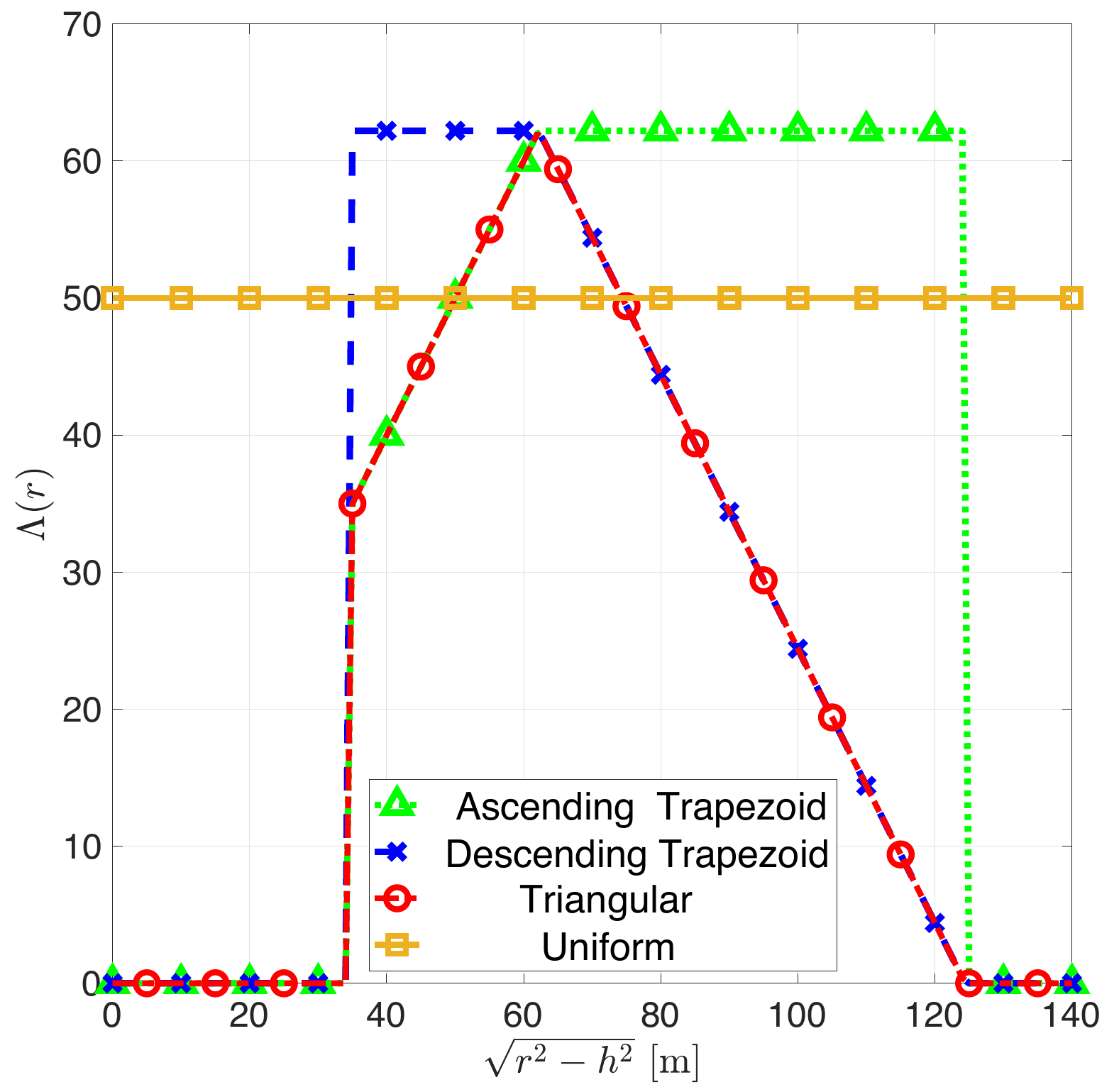}
\caption{Diagram of mass density functions when $h = 20~$m.}
\label{fig1}
\end{minipage}
\hfill
\begin{minipage}[t]{0.52\linewidth}
\centering
\includegraphics[width=0.98\linewidth]{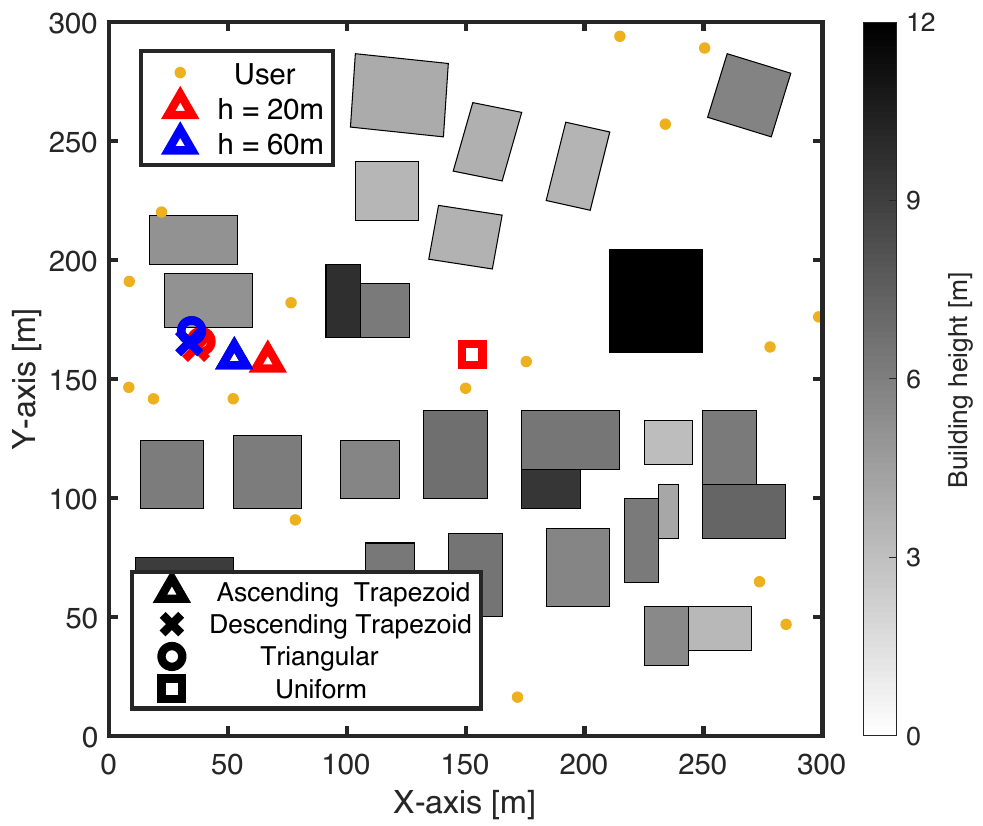}
\caption{A sampling of the BIA.}
\label{fig2}
\end{minipage}
\end{figure}

\subsection{Performance of the Barycenter-Inspired Algorithm}


We start with the mass density function and the BIA. When $h=20~$m, the diagram of four mass density functions $\Lambda(r)$ of the BIA is provided in Fig.\ref{fig1}. For the selection of uniform mass density, it can be regarded that the algorithm will stop at the initialization step. Considering that users who are close to the UAV are likely to be covered, and users who are far away are difficult to be covered by small adjustments to the UAV's position, neither of these users deserves high weight. According to the above analysis, the expressions of the remaining three mass density functions are designed and given in Table~\ref{table2}. $R_{\min}$ and $R_{\max}$ in Table~\ref{table2} are  non-terrain-based boundaries of $C_1$-$C_2$ and $C_2$-$C_3$ in subsection \ref{classification}. In our scenario, they can be observed from Fig.~\ref{fig7} that $R_{\min} = 40~$m and $R_{\max} = 126~$m.

\begin{table}[]
\centering
\caption{Expressions of three mass density functions.} 
\label{table2}
 \renewcommand{\arraystretch}{1.1}
\begin{tabular}{|c|c|c|c|}
\hline
 &
  \begin{tabular}[c]{@{}c@{}}$r \leq \max\{h,R_{\min}\}$\\ or  $R_{\max} < r$\end{tabular} &
  \begin{tabular}[c]{@{}c@{}}$\max\{h,R_{\min}\} < r$\\ and   $r \leq \frac{1}{2}\sqrt{R_{\max}^2 + 3h^2}$ \end{tabular} &
  $\frac{1}{2}\sqrt{R_{\max}^2 + 3h^2} < r \leq R_{\max}$ \\ \hline
Ascending trapezoid  & $\Lambda(r) = 0 $ & $\Lambda(r) = \sqrt{r^2 - h^2}$                   & $\Lambda(r) = \frac{1}{2}\sqrt{R_{\max}^2 - h^2}$         \\ \hline
Descending trapezoid & $\Lambda(r) = 0 $ & $\Lambda(r) = \frac{1}{2}\sqrt{R_{\max}^2 - h^2}$ & $\Lambda(r) = \sqrt{R_{\max}^2 - h^2} - \sqrt{r^2 - h^2}$ \\ \hline
Triangular           & $\Lambda(r) = 0 $ & $\Lambda(r) = \sqrt{r^2 - h^2}$                   & $\Lambda(r) = \sqrt{R_{\max}^2 - h^2} - \sqrt{r^2 - h^2}$ \\ \hline
\end{tabular}
\end{table}


Fig.~\ref{fig2} shows a typical sampling of the BIA. Except for the uniform mass density function, the horizontal positions of the UAV obtained by the remaining three mass density functions are usually close to each other, regardless of the height for $h=20~$m or $h=60~$m. 

\par

\begin{figure*}[ht!]
\begin{minipage}[t]{0.49\linewidth}
\centering
\includegraphics[width=0.98\linewidth]{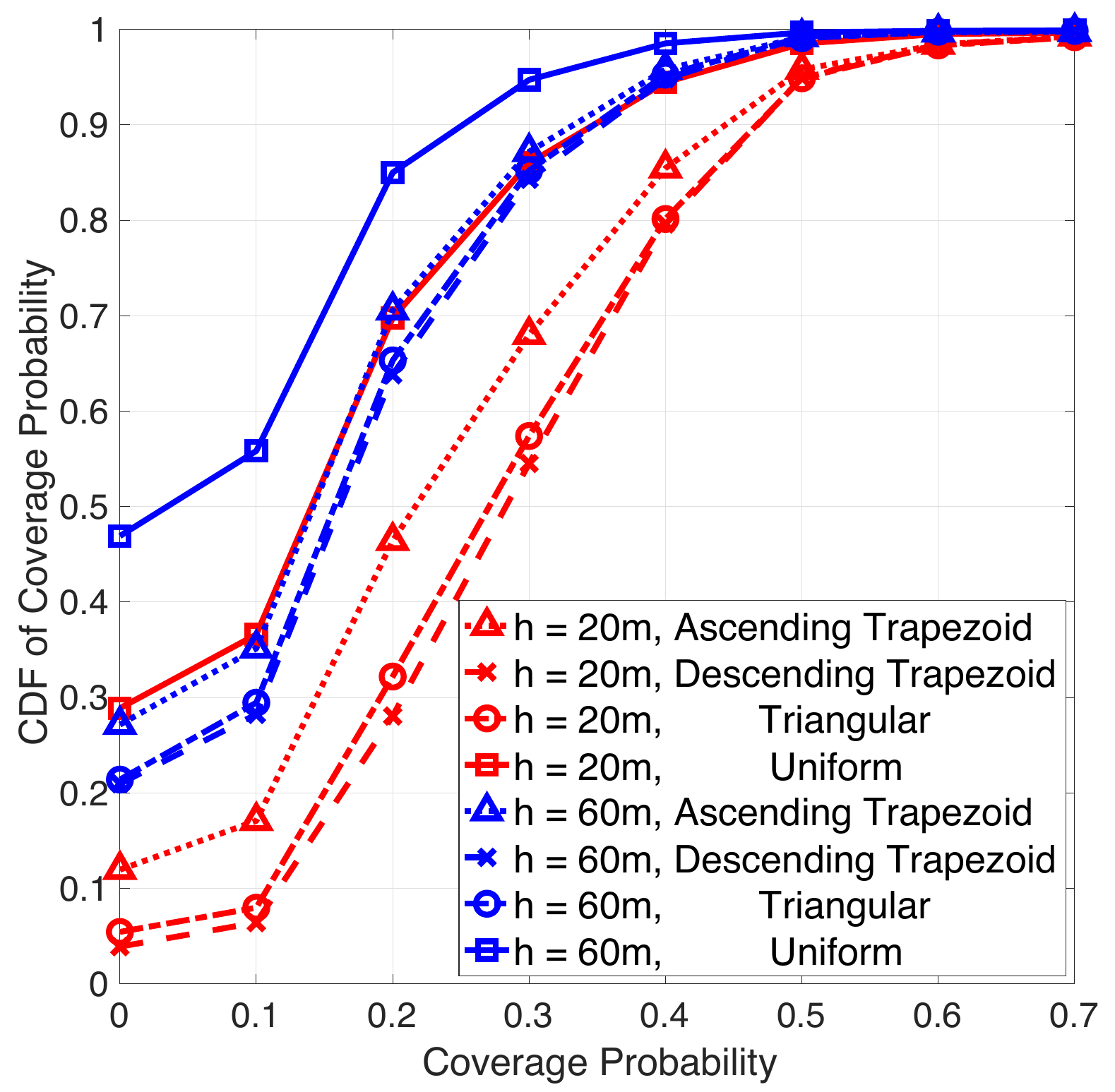}
\caption{Influences of UAV height and mass density function on coverage probability.}
\label{fig3}
\end{minipage}
\hfill
\begin{minipage}[t]{0.49\linewidth}
\centering
\includegraphics[width=0.98\linewidth]{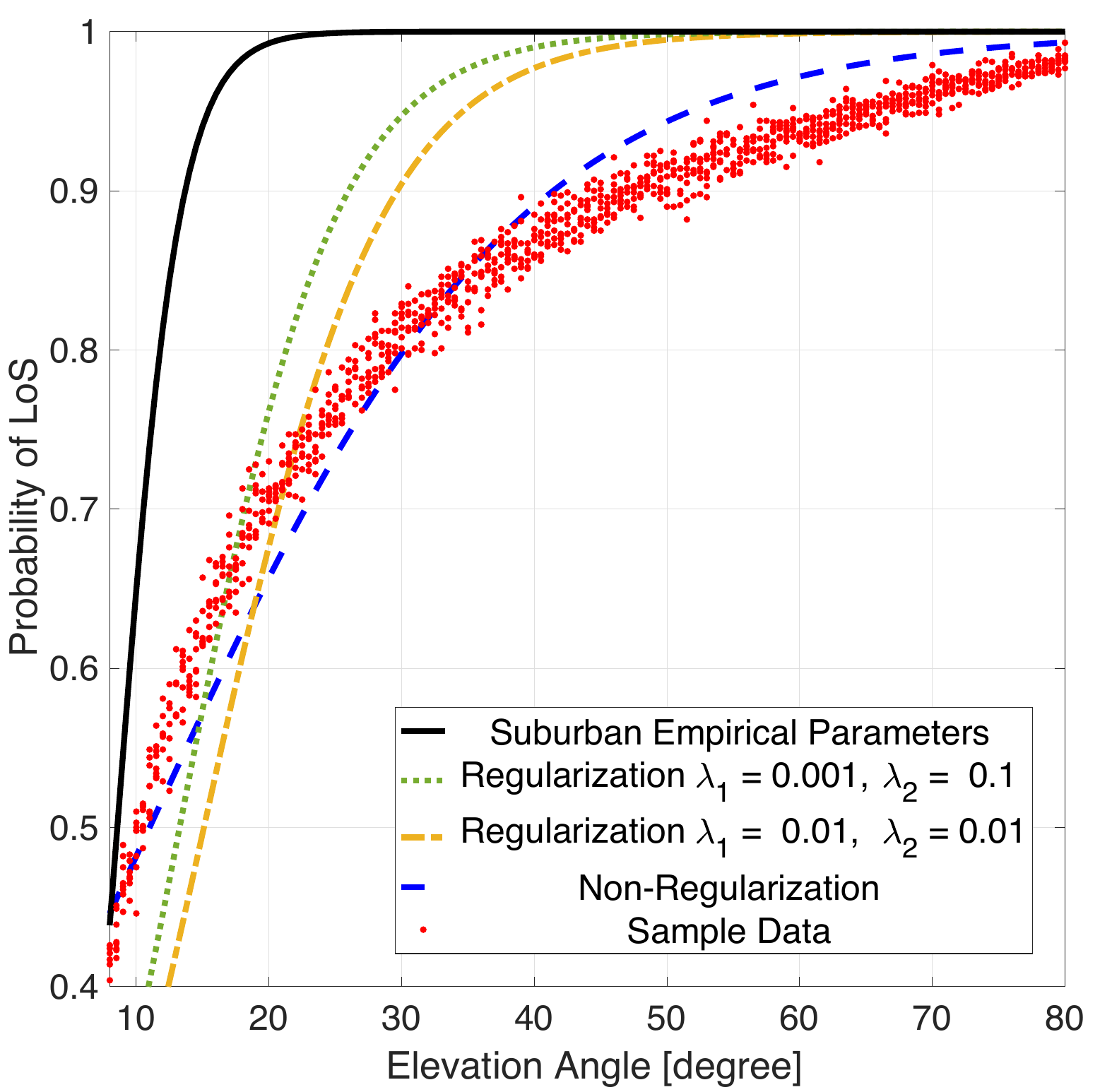}
\caption{Parameter estimation of LoS probability distribution.}
\label{fig4}
\end{minipage}
\end{figure*}

Fig.~\ref{fig3} illustrates that deploying the UAV at relatively low heights and applying non-uniform density functions can effectively improve coverage performance. {\color{black} Note that the coverage probability varies at different locations due to the varying distances to the UAV and the impact of building blockage. After collecting coverage probability samples from different locations across the entire area, we can obtain the cumulative distribution function (CDF) of coverage probability, which is the $y$-axis of Fig.~\ref{fig3} (as well as Fig.~\ref{fig5}, \ref{fig10} and, \ref{fig12}).} When the UAV is deployed at 20~m, the coverage probability is between 10\% and 50\%. When the UAV is deployed at 60~m, the coverage probability is almost always less than 40\%. By comparing CDFs from different density functions at the same height, it can be seen that it is more effective in reducing the weight of users who are farther from the UAV than those who are closer.

\subsection{Performance of Stochastic Channel-Based Positioning Algorithm}\label{simu: algorithm2}
In this subsection, we first provide the numerical results pertaining to the estimation of the LoS probability distribution. Then, we analyze the performance of SCBA.
\par

\begin{table}[H]
\centering
\caption{Comparison of parameters and MSEs under different estimates.} 
\label{table3}
\begin{tabular}{|c|c|c|c|c|c|}
\hline
&$\lambda_{1}$ & $\lambda_{1}$ & $a$ & $b$ & MSE \\ \hline
Empirical parameters & N/A    & N/A         & 4.88  & 0.43    & 0.0220    \\ \hline
Regularization \uppercase\expandafter{\romannumeral1}  & 0.001 & 0.1  & 4.52 & 0.17 & 0.0074 \\ \hline
Regularization \uppercase\expandafter{\romannumeral2}  & 0.01  & 0.01 & 4.76 & 0.15 & 0.0071 \\ \hline
Non-regularization & 0   & 0  & 1.93 & 0.07 & 0.0017 \\ \hline
\end{tabular}
\end{table}

Firstly, the UAV obtains the LoS probability under different elevation angles by random sampling. The probability is estimated as the proportion of LoS links established between the user and the UAV, which is shown as the red dots in Fig.~\ref{fig4}. Based on these data, we present the results of regularized and non-regularized nonlinear regression in Fig.~\ref{fig4}. Empirical parameters of sub-urban environments ($\widehat{a}=4.88$, $\widehat{b}=0.43$, provided by authors in \cite{bor2016efficient}) are used for comparison and regularization.

Table~\ref{table3} provides the parameters $a$ and $b$ and the mean square error (MSE) obtained by different estimations. Both the MSEs in Table~\ref{table3} and the curves in Fig.~\ref{fig4} show that the empirical parameters do not fit well with the LoS probability distribution.
In fact, since only two parameters need to be estimated, relatively accurate results can be obtained by only a small data set with hundreds of samples. Note that if dozens of data are collected for only several elevation angles, smoothing is necessary to eliminate outliers in the data.


\begin{table}[H]
\centering
\caption{Modified functions for the LoS probability distribution and performance comparisons.} 
\label{table4}
\begin{tabular}{|c|c|c|c|c|}
\hline
& Modified expression   & $a$ & \textbf{$b$} & MSE \\ \hline
Sigmoid & $\left(1+{a\exp(-b(\theta-a))}\right)^{-1}$ & 1.93         & 0.07         & 0.0017       \\ \hline
Tanh & $a\frac{ \exp (2b\theta)-1}{ \exp (2b\theta)+1}$ & 0.9463 & 2.7972 & 0.0460 \\ \hline
ReLu    & $\max\{0,a\theta+b\}$      & 0.43         & 0.61         & 980.66       \\ \hline
\end{tabular}
\end{table}

In addition to the modified sigmoid function suggested by authors in \cite{al2014optimal}, we propose expressions of modified tanh and ReLu functions, which are also commonly used functions in machine learning. For a fair comparison, each of these functions contains two undetermined parameters. Table~\ref{table4} compares the expressions and performances of these modified functions. The approximately linear ReLu function is not suitable for the fitting of nonlinear curves, and the modified sigmoid function is the most suitable one for estimating the LoS probability distribution.
\par

\begin{figure*}[htbp]
\begin{minipage}[t]{0.46\linewidth}
\centering
\includegraphics[width=0.98\linewidth]{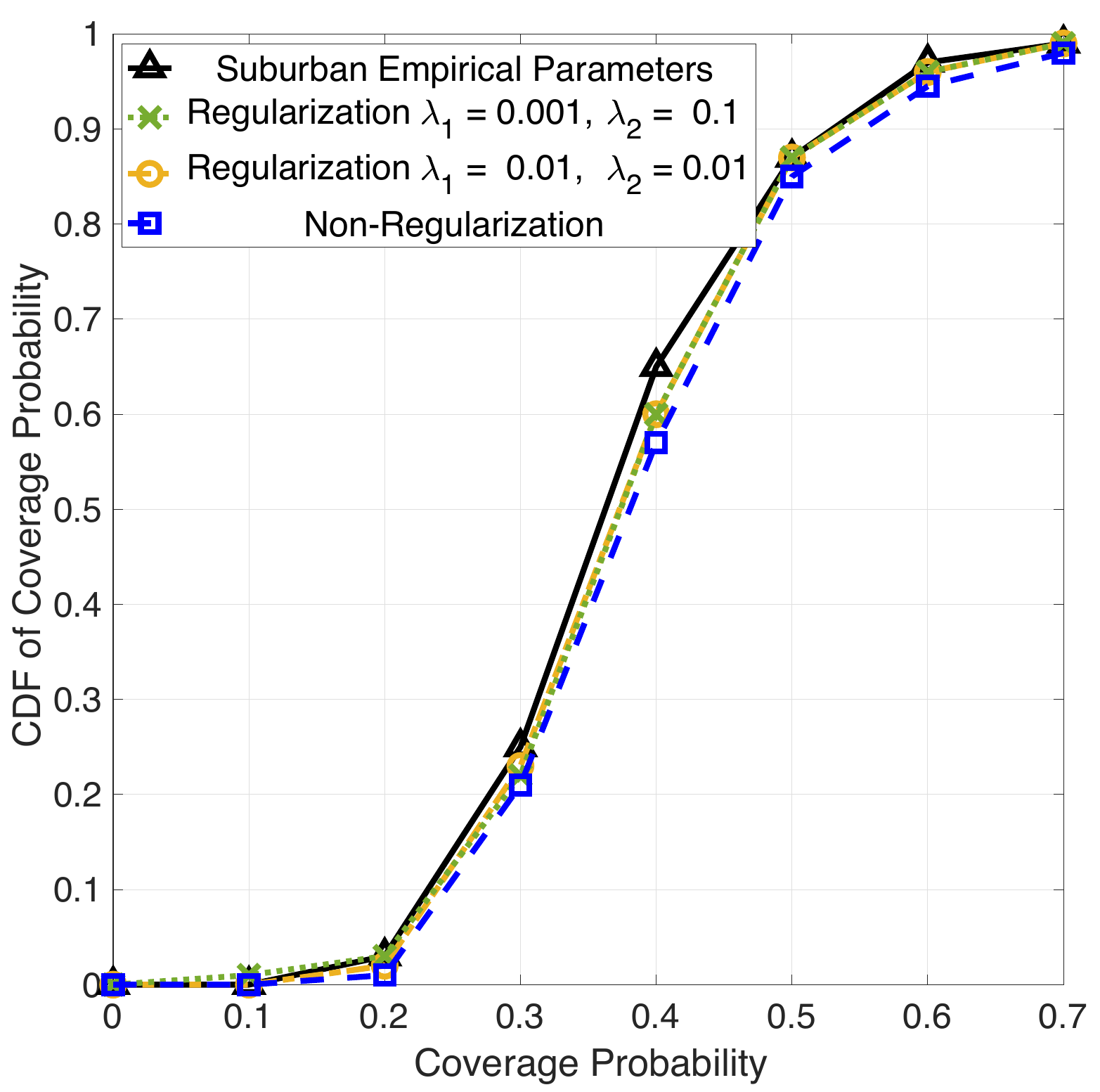}
\caption{SCPA performance under different estimation methods.}
\label{fig5}
\end{minipage}
\hfill
\begin{minipage}[t]{0.52\linewidth}
\centering
\includegraphics[width=0.98\linewidth]{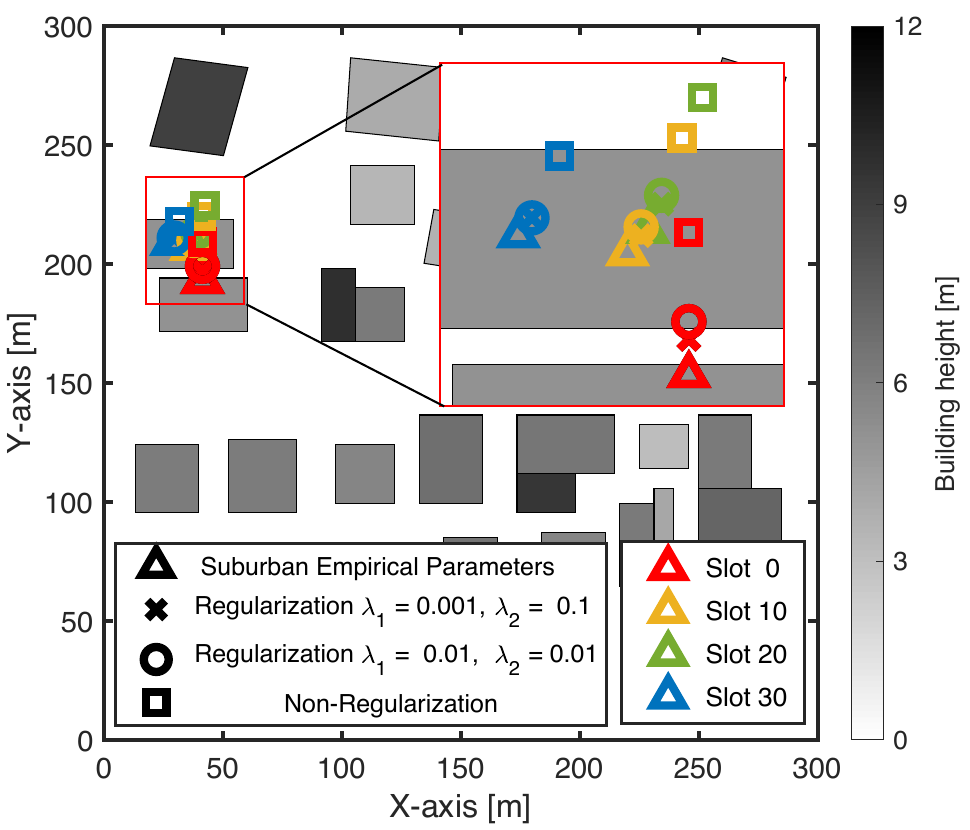}
\caption{{\color{black}A sampling of the UAV positions at different time slots for SCPA.}}
\label{fig6}
\end{minipage}
\end{figure*}

Fig.~\ref{fig5} compares the performances of SCPA under different estimation methods. In general, more accurate LoS probability estimation leads to better performance. The performances of the four methods are similar and significantly better than that of the BIA. 

\par

{\color{black} Fig.~\ref{fig6} shows the positions of the UAV in different time slots obtained by SCPA in a dynamic scenario.} Assume that each user moves less than 3 meters in a random direction (or per second), and users are always located in the outdoor region. The results in Fig.~\ref{fig6} show that regardless of the used estimation method, the positions of the UAV obtained by the SCPA do not change significantly even after 30 slots. It suggests that only a small neighborhood needs to be traversed by brute force.

\subsection{Performance of Algorithms Under Real-Time Search}

\begin{figure*}[htbp]
\begin{minipage}[t]{0.45\linewidth}
\centering
\includegraphics[width=0.98\linewidth]{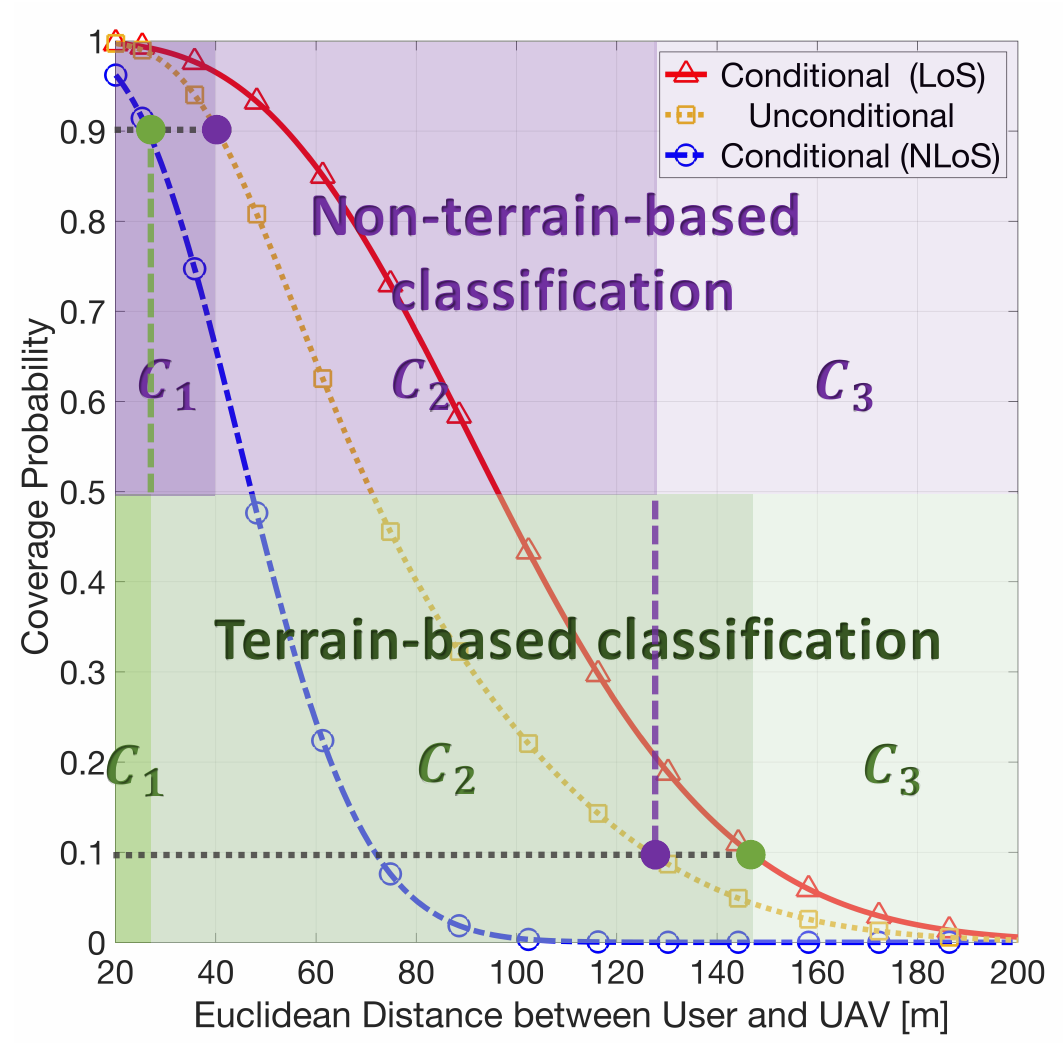}
\caption{User classification from coverage probability perspective when $h=20$~m.}
\label{fig7}
\end{minipage}
\hfill
\begin{minipage}[t]{0.525\linewidth}
\centering
\includegraphics[width=0.98\linewidth]{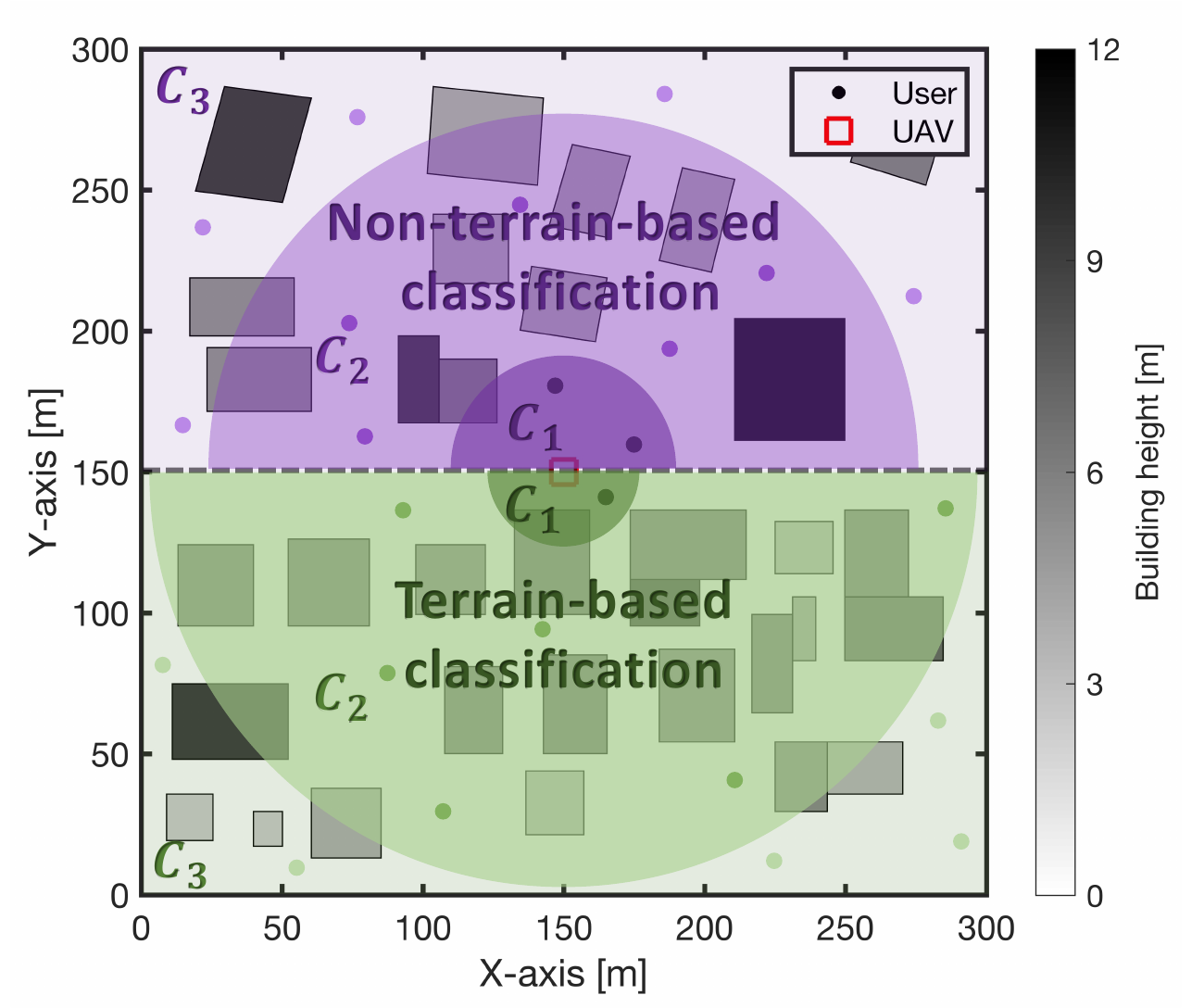}
\caption{User classification from map topology perspective when $h=20$~m.}
\label{fig8}
\end{minipage}
\end{figure*}

As described before, the first step of algorithms under real-time search is user classification. Fig.~\ref{fig7} and Fig.~\ref{fig8} show the results of classification from the coverage probability perspective and map topology perspective, respectively. The height of the UAV is 20~m, while the degree of classification is set to $\varepsilon=0.1$. The details of terrain-based and non-terrain-based classification are described in subsection~\ref{classification}. For the same value of $\varepsilon$, more users will be classified into $C_1$ and $C_3$ in terrain-based classification.

\begin{figure*}[htbp]
\begin{minipage}[t]{0.52\linewidth}
\centering
\includegraphics[width=0.98\linewidth]{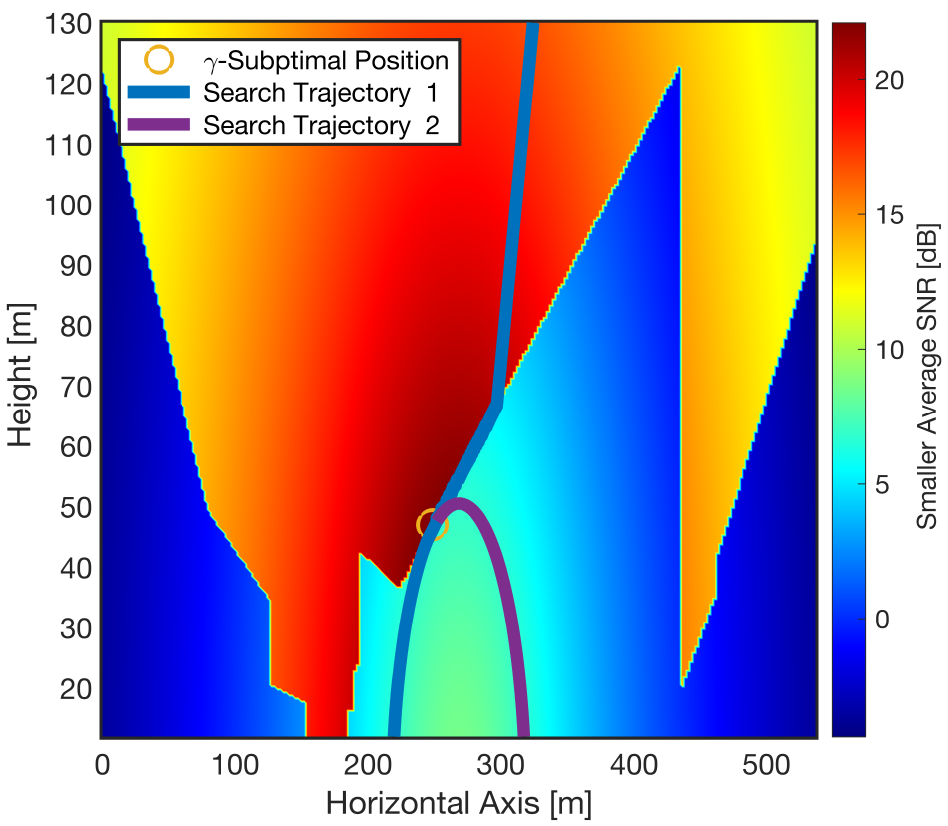}
\caption{An example of Algorithm~\ref{alg1} in a heat map.}
\label{fig9}
\end{minipage}
\hfill
\begin{minipage}[t]{0.455\linewidth}
\centering
\includegraphics[width=0.98\linewidth]{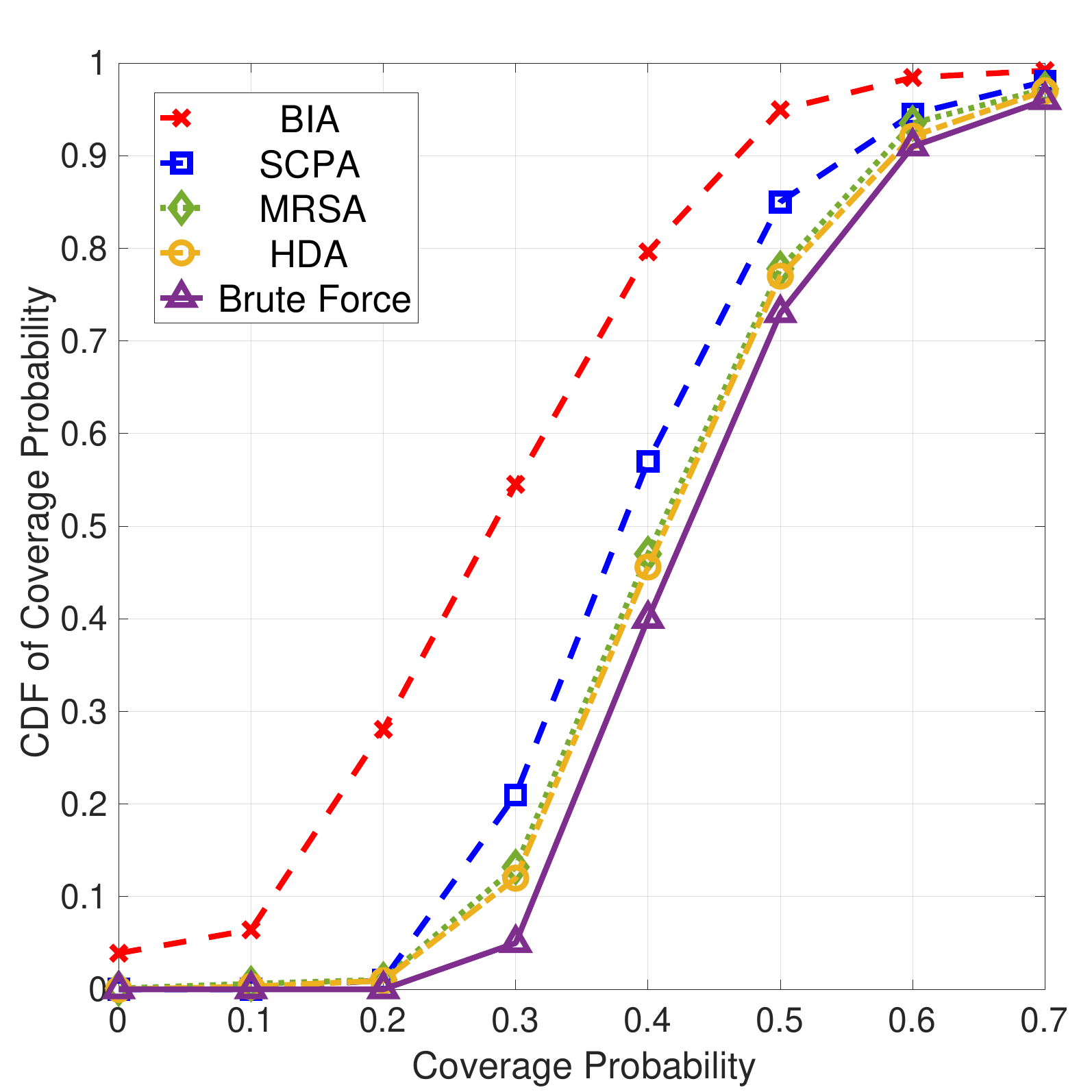}
\caption{Coverage probability performance comparison of all algorithms.}
\label{fig10}
\end{minipage}

\end{figure*}

Fig.~\ref{fig9} provides an example of Algorithm~\ref{alg1} given in subsection~\ref{real-time}. By executing terrain-based classification with $\varepsilon=0.1$, the two farthest users in $C_2$ are selected, and the heat map of smaller average SNR between them is given in Fig.~\ref{fig9}. The heat map corresponds to the middle vertical plane (the $z=0$ plane) of the two users. We can see in Fig.~\ref{fig9} that the UAV starts from the LoS region on the upper right and descends in a straight line in the direction of the steepest gradient. After arriving at the buildings' boundary, the UAV searches along the boundary to find the $\gamma$-suboptimal position. Then, it enters the NLoS area of one of the users and searches the left half branch (blue line) first and then the right half branch (purple line) along the gradient. After confirming that the potential LoS region is not missed, the $\gamma$-suboptimal is finally determined as the output position.

As shown in Fig.~\ref{fig10}, the BIA serves as a comparison since it does not utilize terrain information and only optimizes the horizontal coordinates. The SCPA can further optimize the UAV deployment height based on the former algorithm by extra terrain information collected before networking. Compared with the above two algorithms, the MRSA further improves the coverage performance by avoiding the shadow of buildings. The HDA has a similar coverage performance to the MRSA, and both are close to the upper bound of coverage performance. The upper bound is obtained by an idealized method known as brute force, which traverses every position and leaves the position with the best overall coverage performance. {\color{black} By comparing these five algorithms, we can draw the following conclusions:
    \begin{itemize}
    \item Brute Force versus SCPA: Compared to having partial terrain information (knowing terrain feature parameters), having complete terrain information (precise locations of all buildings) can increase coverage probability by about 6\%. 
    \item SCPA versus BIA:
    Having no terrain information leads to an average loss of 9\% in coverage probability gain compared to SCPA that obtained partial terrain information. 
    \item SCPA versus MRSA: In the case of unrestricted search length, real-time search algorithms can gather more information during the search process than what is contained in the feature parameters, and it brings about a 3\% improvement in coverage probability.
    \item MRSA versus HDA: When search length is not limited, whether mastering the values of feature parameters has little impact on real-time search. 
    \end{itemize}
    }

\begin{figure*}[htbp]
\begin{minipage}[t]{0.49\linewidth}
\centering
\includegraphics[width=0.94\linewidth]{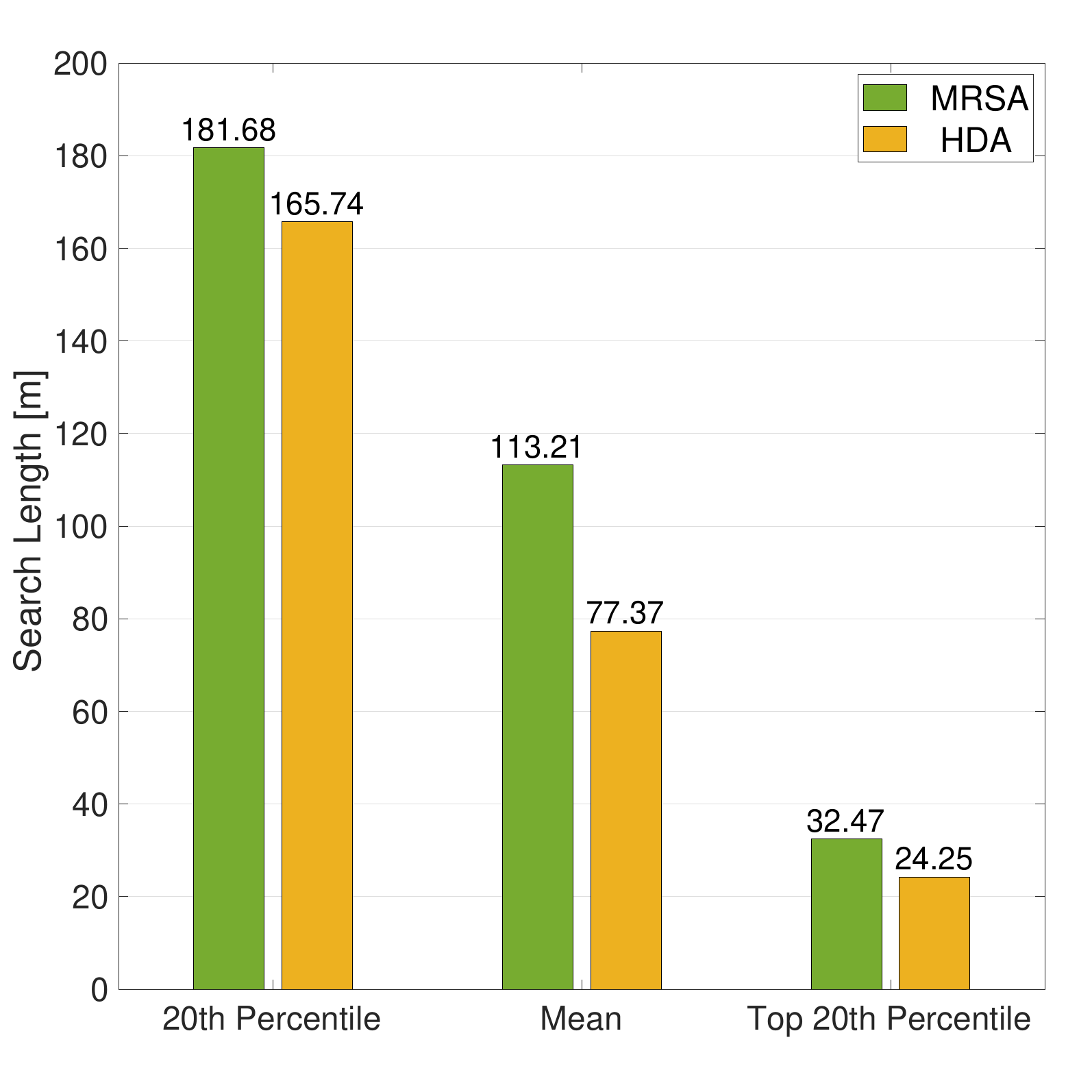}
\caption{Search lengths comparison of MRSA and HDA.}
\label{fig11}
\end{minipage}
\hfill
\begin{minipage}[t]{0.49\linewidth}
\centering
\includegraphics[width=0.94\linewidth]{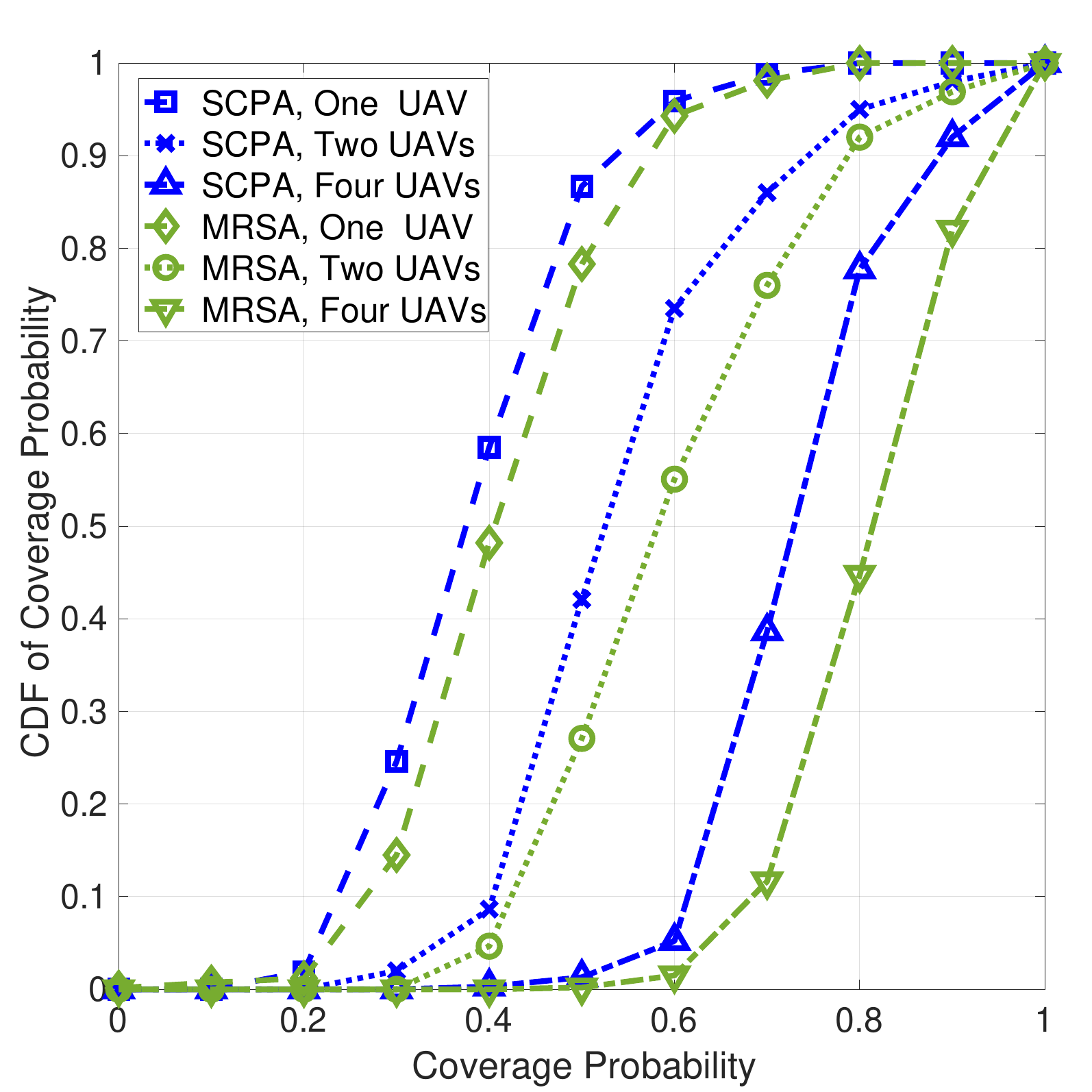}
\caption{Influence of the number of UAVs on coverage probability performance under multiple blockage model.} 
\label{fig12}
\end{minipage}
\end{figure*}

In addition, Fig.~\ref{fig11} shows the boundary of the longest 20\%, the mean and the boundary of the shortest 20\% of the samples of search length for the above two algorithms. Compared to the MRSA,  the HDA shortens the search distance with the help of the collected terrain information.


\subsection{Extended Scenarios}

In this subsection, we investigate a more sophisticated and realistic scenario in which numerous UAVs serve a region simultaneously, and the effect of multiple blockages of the buildings is considered. The following describes the impact of these two factors on algorithm operation.
\begin{itemize}
    \item Multiple UAVs Deployment:
    Firstly, the entire area is divided into sub-regions with equal area. The number of sub-regions is equal to the number of UAVs, and the sub-regions correspond to UAVs one by one. When the algorithm is excused to determine the networking position, the UAV only considers the users in its corresponding sub-region. Notice that the user will select the UAV providing the maximum received power to be associated with, thus, the user might connect to a UAV in another sub-region. 
    \item Modeling of Multiple Blockages: When the link between the user and the UAV is blocked twice or more, we assume the signal from this UAV cannot be received by the user.
\end{itemize}

Fig. \ref{fig12} shows the performance of the SCPA and the MRSA while the number of UAVs is set to one, two, and four. Compared with the results in Fig. \ref{fig10}, the performance of a single UAV is slightly worse than before because of multiple blockages, but being blocked more than once is relatively low since buildings are sparse in our scenario. Furthermore, no matter how many UAVs are deployed, the SCPA (blue lines) consistently outperforms the MRSA (green lines). 
In addition, as the number of UAVs increases, the performance improves. This is due to the fact that the average distance between users and UAVs decreases and the LoS probability increases.

\subsection{Summary and Algorithm Comparison} 
We investigated the impact of the UAV height and the selection of mass density function on the coverage performance of BIA. From the obtained results, we concluded that it is more effective to reduce the weight assigned to the users who are farther from the UAV than those who are closer. We showed the estimation parameters $a$ and $b$ in the LoS probability model under different regularization parameters. We recall that $a$ and $b$ are obtained by ridge regression, which contains regularization parameters. The results have shown that the empirical parameters do not fit well with the LoS probability distribution.

Therefore, we tested the performance of SCPA with different values of $a$ and $b$ under different regularization parameters, and showed that in general, more accurate LoS probability estimation provided better, albeit minor, improvement in the performance.
We also showed that SCPA performs significantly better than BIA.
Then, we compared the coverage performance of four algorithms with the upper bound, that is, the brute force method.
The BIA serves as a lower bound since it does not utilize terrain information and only optimizes the horizontal coordinates.
On the other hand, the SCPA can further optimize the UAV deployment height based on the former algorithm by extra terrain information collected before networking, and therefore, has better performance than the BIA.
Compared with the above two algorithms, the MRSA further improves the coverage performance by avoiding the shadow of buildings. The HDA has a similar coverage performance to the MRSA, and both are close to the upper bound of coverage performance. 
We compared the search lengths of HDA and MRSA since they are two real-time search methods, the difference between them is that the former has offline terrain information and the latter does not.
Compared to the MRSA, the HDA shortens the search distance with the help of the collected terrain information.

Finally, we extended the scenario using only one UAV to a general scenario using multiple UAVs.
We also extended the original blockage model from using only LoS and NLoS into a multiple blockage model, which includes no blockage, blockage once, and blockage twice or more than twice.
We showed that as the number of UAVs increases, the performance improves.
However, the performance of a single UAV is slightly worse compared to the scenario using only LoS and NLoS. 
This minor decrease in performance is caused by the multiple blockages, however, the probability of being blocked more than once is relatively low since buildings are sparse in the considered scenario.

\section{Conclusion}
In this paper, two methods were proposed to improve the coverage of the UAV network in combination with the actual terrain information before and during networking. Before networking, the UAV can estimate the probability that UAV and users are blocked by buildings at different elevation angles through offline information collection. Furthermore, we derived the analytical expression of coverage probability and classified users accordingly. Classification results can simplify the optimization problem and narrow the range of users that UAVs need to focus on. During networking, we analyzed the optimality of UAV networking position in simple scenes, and further designs the real-time search method. Furthermore, we designed four algorithms according to whether the above two UAV deployment methods are used or not. Finally, in order to obtain more feasible experimental results, multiple UAV deployments and multiple blockages are considered as further extensions.

\appendices

\section{Proof of Theorem \ref{P_C}}\label{app:P_C}
According to the definition in (\ref{PC_definition}), the coverage probability for a user whose distance from the UAV is $r_k$ can be expressed as,
\begin{equation}
\begin{split}
    &P_{C,k} \left( h , r_k \right) = P_{\rm{LoS}} \left( h , r_k \right) \mathbb{P} \left[ \frac{S_{{\rm{LoS}},k}\left(r_k\right)}{\sigma^2}>\gamma \right] \\
    & + P_{\rm{NLoS}} \left( h , r_k \right) \mathbb{P} \left[ \frac{S_{{\rm{NLoS}},k}\left(r_k\right)}{\sigma^2}>\gamma \right] \\
    &\overset{(a)}{=} P_{\rm{LoS}} \left( h , r_k \right) \frac{\Gamma_u \left( m_{\rm{LoS}}, m_{\rm{LoS}} \eta_{\rm{LoS}}^{-1} \zeta^{-1} \gamma \, r_k^{\alpha_{\rm{LoS}}} \sigma^2  \right) }{\Gamma\left( m_{\rm{LoS}} \right)} \\
    & + P_{\rm{NLoS}} \left( h , r_k \right) \frac{\Gamma_u \left( m_{\rm{NLoS}}, m_{\rm{NLoS}} \eta_{\rm{NLoS}}^{-1} \zeta^{-1} \gamma \, r_k^{\alpha_{\rm{NLoS}}} \sigma^2  \right) }{\Gamma\left( m_{\rm{NLoS}} \right)} \\
    &\overset{(b)}{=} P_{\rm{LoS}} \left( h , r_k \right) \exp\left(- \mu_{\rm{LoS}} \left( r_k \right) \right) \sum_{n=0}^{m_{\rm{LoS}}-1} \frac{\mu_{\rm{LoS}} \left( r_k \right)^n}{n!}\\
    & + P_{\rm{NLoS}} \left( h , r_k \right) \exp\left(- \mu_{\rm{NLoS}} \left( r_k \right) \right) \sum_{n=0}^{m_{\rm{NLoS}}-1} \frac{\mu_{\rm{NLoS}} \left( r_k \right)^n}{n!},
\end{split}
\end{equation}
where $(a)$ is from the complementary cumulative distribution function (CCDF) of the Gamma function $\overline{F}_G \left( g \right) = \frac{\Gamma_u\left(m,mg\right)}{\Gamma\left( m \right)}$, and $\Gamma_u\left(m,mg\right) = \int_{mg}^{\infty} t^{m-1} e^{-t} dt$ is the upper incomplete Gamma function. $(b)$ is derived by the definition $\frac{\Gamma_u\left(m,mg\right)}{\Gamma\left( m \right)} = \exp(-g) \sum_{n=0}^{m-1} \frac{g^n}{n!}$ and substituting $\mu_Q \left( r_k \right) = \eta_{Q}^{-1} \zeta ^{-1} \gamma \, r_k^{\alpha_Q} \sigma^2, \ Q=\left\{\rm{LoS},\rm{NLoS}\right\}$ into  the expression in $(a)$.

\section{Proof of Proposition~\ref{proposition1}}\label{app:proposition1}
In scenario~\uppercase\expandafter{\romannumeral1}, the height of the optimal position is $h_{\min}$ since there are no buildings. Therefore, only the horizontal position of the UAV remains to be optimized. 
\par
When a UAV serves only one user, the optimal horizontal position is the user's horizontal position. In other words, the UAV hovers over the user. 
\par
When covering two users who are close to each other, the optimal horizontal position of the UAV might be the midpoint of the two users. When two users are far apart, there are two optimal locations where there is symmetry about the midpoint of the two users. When two users are infinitely far apart, the UAV hovers over one of them. In either case, the optimal location is on the line segment connecting the two users and is therefore linearly optimal.
\par
When the UAV provides coverage to more than three users, we simplify the optimization objectives as follows:
\begin{equation}
\begin{split}
     & P_C \left(h_{\min};r_1,...,r_K\right) = \frac{1}{K} \sum_{k=1}^K \mathbb{P} \left[ \frac{S_{Q,k}\left(r_k\right)}{\sigma^2}>\gamma \right]\\
     & \overset{(a)}{\approx} \frac{1}{K} \sum_{k=1}^K v \sqrt{r_k^2 + h_{\min}^2} + w \overset{(b)}{\approx} \frac{1}{K} \sum_{k=1}^K v r_k + w,
\end{split}
\end{equation}
where $(a)$ is obtained by approximating the coverage probability as a linear function of the Euclidean distance between the UAV and the users. Through classification, the coverage probabilities of users in $C_2$ locate in the approximately linear region of the covering probability function, so this assumption is reasonable. $(b)$ is based on the fact that $r_k$ is often much larger than $h_{\min}$ for $C_2$ users. 
\par
When $K=3$, the optimal position is called the Fermat-Weber point. When $K \geq 4$, the authors in \cite{carmi2005fermat} proposed a method to approach the optimal position based on the above approximation. Since the number of iterations is finite, the search trajectory is linear. However, it requires that the polygon connected by users is a convex polygon. In addition, this linear search trajectory cannot approach with an arbitrarily small precision, therefore, it is not linear optimal.

\section{Proof of Proposition~\ref{proposition2}}\label{app:proposition2}
The formula (\ref{S_Q}) indicates that for a given UAV position, the set of positions on the ground satisfying $\overline{\rm{SNR}}_Q(r) \geq \gamma$ is a circle. 
Therefore, if all the users are located inside a circle or on its boundary,
then the center of that circle is one of the suboptimal solutions. This problem is called the circle covering problem. In \cite{hearn1982Efficient}, the author provides a minimum circle covering algorithm. The algorithm keeps two points (users) on the boundary of the covering circle, repeatedly shrink the circle and replaces the boundary points (users) until the optimal covering circle is found. Through this algorithm, the UAV can find the suboptimal location without searching, thus, a $\gamma$-suboptimal strategy $\mathscr{L}$ exists.

\section{Proof of Proposition~\ref{proposition3}}\label{app:proposition3}
Before starting the proof, we denote the UAV location  that is not blocked by the building for both users as LoS position, otherwise NLoS position. Under the condition that $2\rho^* < d$, the proof of Proposition~\ref{proposition3} is divided into the following four steps:
\begin{itemize}
    \item \textbf{Step 1:} The $\gamma$-suboptimal position is located on the boundaries caused by the shadows of the buildings, or on the plane of height $h_{\min}$.
    \item \textbf{Step 2:} In most cases, the $\gamma$-suboptimal position is located in the plane $z=0$.
    \item \textbf{Step 3:} The output $(\rho^* , \theta^* , 0)$ is the $\gamma$-suboptimal position.
    \item \textbf{Step 4:} The search length of Algorithm~\ref{alg1} is linear with respect to the height of UAV $h$ and the distance between users $d$.
\end{itemize}

\textbf{Step 1:} For LoS positions, the average SNR of both users is enhanced with the decrease of the UAV height. For NLoS positions, the average SNR of the users with weak signal power is enhanced with the decrease of the UAV height. Due to the minimum height restrictions, the $\gamma$-suboptimal position is located on the boundaries caused by the shadows of the buildings (corresponding to the LoS positions), or on the plane of height $h_{\min}$ (corresponding to the NLoS positions).

\textbf{Step 2:} The right side of the Fig.~\ref{fig:big} shows an inclined plane that can be uniquely determined by three points on the plane, that is, the coordinates of two users $\left(0,0,\pm \frac{d}{2}\right)$ and an arbitrary point $\left(\rho_0, \theta_0, 0\right)$, where $\theta_0$ is the inclination between the plane and the ground. Using basic geometry, we can draw the following conclusion: for an arbitrarily small $\Delta \rho \rightarrow 0$, if $\left(\rho_0^*+\Delta \rho, \theta_0,0 \right)$ is a LoS position and $\left(\rho_0^*-\Delta \rho, \theta_0,0 \right)$ is a NLoS position ($\rho_0^* \sin\theta_0 > h_{\min}$), the smaller average SNR of any LoS position in the plane with inclination of $\theta_0$ is less than $\overline{\rm{SNR}}_{\rm{LoS}}\left( \max \left\{ \frac{2 \rho_0^* d}{\sqrt{\left(2\rho_0^*\right)^2 + d^2}}, \sqrt{\left(\rho_0^*\right)^2 + \frac{d^2}{4}} \right\} \right)$. When $2\rho_0^* < d$, $\left(\rho_0^*, \theta_0,0 \right)$ is the position with the maximum smaller average SNR, denoted as the inclined-suboptimal position.  Because in most cases, $2\rho_0^* < d$ is satisfied (in simulation, $d \approx 120$m and $\rho^* \approx 20$m usually), thus, the suboptimal position is located in the plane $z=0$.

\textbf{Step 3:} We extend the trajectory to the left and right of the straight line $h=h_{\min}$ in the plane $z=0$. According to the monotonically decreasing value of $\rho$ in algorithm 1, $(\rho^*, \theta^*, 0)$ has the shortest distance to the users among the points on the search trajectory.  It is straightforward to prove that any point on the extended trajectory of Algorithm~\ref{alg1} is less than or equal to the distance from the corresponding inclined-suboptimal position to the user. Since $(\rho^*, \theta^*, 0)$ is a LoS position, and has the shortest distance to the users among the inclined-suboptimal positions, $(\rho^*, \theta^*, 0)$ is the $\gamma$-suboptimal position.

\textbf{Step 4:} Since one lower bound of the search trajectory is $2\arccos \left(\frac{h_{\min}}{\rho^*}\right) \sqrt{\left(\rho^*\right)^2 + \frac{d^2}{4}}$ and one upper bound is $2\arccos \left(\frac{h_{\min}}{\rho_{\max}}\right) \sqrt{\left(\rho_{\max}\right)^2 + \frac{d^2}{4}}$, where $\rho_{\max}$ is $\rho_u$ after step (6) in Algorithm~\ref{alg1}, thus the trajectory is linear. Finally, note that although we have shown that the $\gamma$-subpotimal position might be located on the boundary and in the plane $z=0$ in most cases, traversing along the boundary does not guarantee linearity. Because boundaries can be complex and finding an upper bound of the search trajectory is not guaranteed.

\bibliographystyle{IEEEtran}
\bibliography{references}

\end{document}